\documentclass[pdftex,sigconf,nonacm]{acmart}

\usepackage{subcaption}
\usepackage{tikz}
\usepackage{mdframed}
\usepackage[normalem]{ulem}
\usepackage{graphicx}
\usepackage{pslatex}          

\usepackage{xspace}

\long\def\ignore#1{}
\sloppypar

\newcommand*\circled[1]{\tikz[baseline=(char.base)]{
  \node[shape=circle,draw,fill=black,text=white,font=\bf,inner sep=0.5pt] (char)
  {\scriptsize#1};
}}

\newcommand{\ApproxSign}{\raise.17ex\hbox{$\scriptstyle\sim$}}

\newcommand{\putsec}[2]{\vspace{-0.0in}\section{#2}\label{sec:#1}\vspace{-0.0in}}
\newcommand{\putssec}[2]{\vspace{-0.0in}\subsection{#2}\label{ssec:#1}\vspace{-0.0in}}
\newcommand{\putsssec}[2]{\vspace{-0.0in}\subsubsection{#2}\label{sssec:#1}\vspace{-0.0in}}

\newcommand{\tabref}[1]{Table~\ref{#1}}
\newcommand{\figref}[1]{Figure~\ref{#1}}
\newcommand{\secref}[1]{Section~\ref{sec:#1}}
\newcommand{\ssecref}[1]{Section~\ref{ssec:#1}}
\newcommand{\sssecref}[1]{Section~\ref{sssec:#1}}

\newcommand{\COMMENT}[1]{#1}

\newcommand{\revise}[1]{{\color{black} #1}} 

\newcommand{\remove}[1]{\COMMENT{{\color{red}\sout{#1}}}}
\renewcommand{\remove}[1]{}

\mdfdefinestyle{verdictbox}
{
    linewidth=0.7pt,
    innertopmargin=2pt,
    innerbottommargin=2pt,
    innerrightmargin=2pt,
    innerleftmargin=2pt,
    leftmargin=0pt,
    rightmargin=0pt,
    nobreak=false,
    splitbottomskip=0pt,
}

\newcommand{\squishlist}{
   \begin{list}{$\bullet$}
    { \setlength{\itemsep}{0pt}      \setlength{\parsep}{0pt}
      \setlength{\topsep}{3pt}       \setlength{\partopsep}{0pt}
      \setlength{\listparindent}{-2pt}
      \setlength{\itemindent}{-5pt}
      \setlength{\leftmargin}{1em} \setlength{\labelwidth}{0em}
      \setlength{\labelsep}{0.5em} } }

\newcommand{\squishend}{
    \end{list}  }

\newcommand{\squishlisttwo}{
   \begin{list}{$\bullet$}
    { \setlength{\itemsep}{0pt}    \setlength{\parsep}{0pt}
      \setlength{\topsep}{0pt}     \setlength{\partopsep}{0pt}
      \setlength{\leftmargin}{2em} \setlength{\labelwidth}{1.5em}
      \setlength{\labelsep}{0.5em} } }

\newcommand{\PNAME}{\mbox{Pimacolaba}\xspace}
\newcommand{\PNAMEBOLD}{\mbox{\textbf{Pimacolaba}}\xspace}

\newcommand{\base}{\mbox{\textit{pim-base}}\xspace}
\newcommand{\colab}{\mbox{\textit{pim-colab}}\xspace}

\AtBeginDocument{%
  \providecommand\BibTeX{{%
    \normalfont B\kern-0.5em{\scshape i\kern-0.25em b}\kern-0.8em\TeX}}}

\setcopyright{acmcopyright}
\copyrightyear{2018}
\acmYear{2018}
\acmDOI{XXXXXXX.XXXXXXX}

\acmConference[Conference acronym 'XX]{Make sure to enter the correct
  conference title from your rights confirmation emai}{June 03--05,
  2018}{Woodstock, NY}
\acmBooktitle{Woodstock '18: ACM Symposium on Neural Gaze Detection,
 June 03--05, 2018, Woodstock, NY} 
\acmPrice{15.00}
\acmISBN{978-1-4503-XXXX-X/18/06}

\begin{document}

\title[Collaborative Acceleration for FFT on Commercial Processing-In-Memory Architectures]{Collaborative Acceleration for FFT on \\Commercial Processing-In-Memory Architectures}

\author{Mohamed Assem Ibrahim}
\email{mohamed1.ibrahim@amd.com}
\affiliation{%
  \country{Advanced Micro Devices, Inc.}
}

\author{Shaizeen Aga}
\email{shaizeen.aga@amd.com}
\affiliation{%
  \country{Advanced Micro Devices, Inc.}
}


\begin{abstract}
This paper evaluates the efficacy of recent commercial processing-in-memory (PIM) solutions to accelerate fast Fourier transform (FFT), an important primitive across several domains. Specifically, we observe that efficient implementations of FFT on modern GPUs are memory bandwidth bound. As such, the memory bandwidth boost availed by commercial PIM solutions makes a case for PIM to accelerate FFT.
To this end, we first deduce a mapping of FFT computation to a strawman PIM architecture representative of recent commercial designs. We observe that even with careful data mapping, PIM is not effective in accelerating FFT. To address this, we make a case for collaborative acceleration of FFT with PIM and GPU. Further, we propose software and hardware innovations which lower PIM operations necessary for a given FFT. Overall, our optimized PIM FFT mapping, termed \PNAMEBOLD, delivers performance and data movement savings of up to 1.38$\times$ and 2.76$\times$, respectively, over a range of FFT sizes.
\end{abstract}



\maketitle

\putsec{s01}{Introduction}

Discrete Fourier transform (DFT) is an important primitive across several domains of import (molecular dynamics, computational chemistry, data analysis and more) and it forms the key building block of important computations (e.g., solving partial differential equations). 
Consequently, efficient implementations of DFTs, specifically of fast Fourier transform (FFT)~\cite{Cooley1965}, have received considerable attention
\revise{from} widely deployed accelerators such as GPUs, which power seven out of ten fastest supercomputers~\cite{top500Nov22}. 

We observe in this work that efficient implementations of FFTs on GPUs are memory bandwidth bound and as such could benefit from techniques which avail higher memory bandwidth than available at the GPU. As such, we evaluate the efficacy of recent commercial processing-in-memory (PIM) solutions~\cite{hynixPIM,mi100PIM}, which avail memory bandwidth boost 
\revise{(potentially up to 12$\times$ as projected in \ssecref{motiv_pim_mem_bw})}
over GPUs by pushing compute to near-memory compute units to accelerate FFT. 
To this end, we begin with deducing compute orchestration and data mapping necessary to map FFT computation to in-memory compute units. We observe here that even with careful compute orchestration and considerable care to map data to memory structures (DRAM banks) appropriately as necessary for PIM, (except for small sizes) PIM leads to considerable slowdown vis-a-vis a GPU (\revise{average} slowdown of about 52\%). 

To tackle this, we make a case for collaborative acceleration of FFT with PIM. That is, we propose that for a given FFT size, harnessing PIM for a judicious portion of the computation is a superior strategy than for the entire computation. To achieve this, we augment existing FFT decomposition mechanism~\cite{msft_fft_paper,msft_fft_paper_sc}, which decomposes a given FFT into component computations all mapped to GPU, to also map some resultant components to PIM  (using our FFT PIM routines). By carefully choosing the portion of computation that is mapped to PIM, we can harness PIM for FFT acceleration (max speedup of about 1.07$\times$). As we will show, such a collaborative acceleration strategy, beyond performance, also has the effect of lowering data movement (up to 2.76$\times$) which has the potential to translate to energy savings. 

Next, we further analyze the operation profile of our proposed collaborative FFT PIM mapping and observe that the majority of the operations are compute commands to in-memory \revise{compute} units. As such, we propose two innovations which help lower PIM compute operations necessary for a given FFT. First, we observe that the butterfly computation~\cite{autofftsc19} in FFT, which is the key building block of a FFT computation, can be decomposed to 
\revise{fewer}
PIM compute operations in certain scenarios (software optimization). Second, we identify a common computation pattern in the butterfly computation and propose simple extension to in-memory compute units which accelerates this pattern (hardware optimization). Our proposed software and hardware optimizations along with our collaborative acceleration strategy, all of which together we term \PNAMEBOLD, deliver performance of up to 1.38x over a range of FFT sizes.

Finally, we observe that as accelerators and processors alike are coupled with memory, solutions like PIM can serve as augmentations over and above existing FFT acceleration solutions that only harness processor-side optimizations. As such, our work complements the rich spectrum of existing (and potentially future) efforts which aim to accelerate the important primitive of FFT. 

We summarize the key innovations and contributions in this work below.
\begin{itemize}
    \item 
    This is the first work to evaluate the efficacy of the emerging commercial processing-in-memory (PIM) solutions, which avail memory bandwidth boost, to accelerate FFT. To this end, we deduce a mapping of FFT computation to a strawman PIM architecture representative of recent commercial PIM designs.
    \item 
    Using the above FFT mapping, we show that even with careful data mapping and compute orchestration, (except for small sizes), PIM leads to considerable slowdown vis-a-vis a GPU (\revise{average} slowdown of about 52\%). 
    \item To tackle the above challenge, we propose collaborative acceleration of FFT with PIM. That is, we augment existing FFT decomposition mechanism, which decomposes a given FFT into component computations all mapped to GPU, to also map some resultant components to PIM. 
    We observe that such a strategy helps achieve max speedup of about 1.07$\times$ and data movement savings of up to 2.76$\times$, a key challenge in today's systems.
    \item Next, we further analyze our collaborative FFT PIM mapping and propose innovations (augmentation to in-memory compute units and software optimization) which lower PIM operations necessary for a given FFT. Our resultant PIM FFT mapping, which we term, \PNAMEBOLD, delivers performance of up to 1.38x over a range of FFT sizes.  
    \item We believe our work presents a complimentary acceleration strategy for an important primitive like FFT that plays well with spectrum of existing (and potentially future) acceleration solutions for FFT. 
\end{itemize}
\putsec{bckg}{Background} 

In this section, we first provide a brief background of fast Fourier transform (FFT) algorithm. Next, we discuss the key characteristics of efficient implementations of FFT on GPUs, 
which power several top supercomputers tackling key HPC problems~\cite{top500Nov22}.
Finally, we provide a background on commercial processing-in-memory (PIM) designs which we study in this work. 

\putssec{bckg_fft}{Fast Fourier Transform (FFT)}

\begin{figure}[t]
    \centering
    \includegraphics[width=\linewidth]{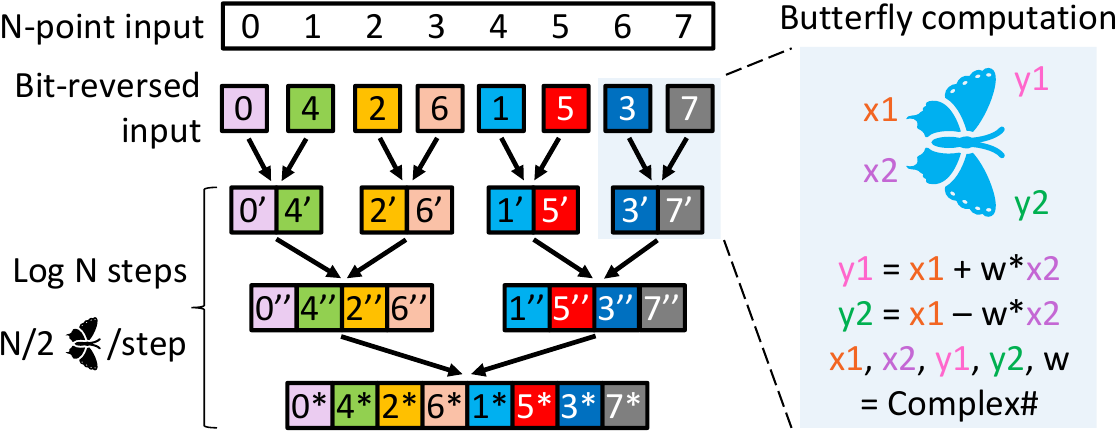}
    \caption{FFT algorithm for size N and butterfly computation.}
    \label{fig:bg_fft}
    \vspace{-\baselineskip}
\end{figure}

The discrete Fourier transform (DFT) transforms a representation of a function in time-domain to its representation in frequency-domain. DFTs are an important primitive across several domains of import (e.g., molecular dynamics, computational chemistry and more). We focus in this work on an efficient method to calculate DFT, namely, fast Fourier transform (FFT) and more specifically on the Cooley-Tukey algorithm~\cite{Cooley1965}, 
a
widely used and efficient algorithm for FFT. Further, we also focus on complex DFT, which transforms two N point time domain functions into two N point frequency domain functions. We discuss 
other forms of FFT in \secref{discuss}.

\figref{fig:bg_fft} shows a simplified view of this efficient FFT algorithm. The algorithm takes as input an array of $N$ samples (complex numbers) in a signal (termed as FFT size henceforth) and sorts them in bit-reversed order. The key building block of FFT algorithm is the butterfly computation. Each butterfly computation takes, as its inputs, two complex numbers $x_1$ and $x_2$ (which are two points in the input) and $\omega$, which is another complex number called twiddle factor. As depicted in \figref{fig:bg_fft} (right), the butterfly computation involves complex number multiplication and addition to produce as outputs two other complex numbers $y_1$ and $y_2$. The FFT algorithm comprises $log(N)$ steps each involving $N/2$ butterfly computations as depicted in \figref{fig:bg_fft} (left). Note that the input to each step is the output of the previous step as depicted. 

\putssec{bckg_efficient_gpu}{Efficient FFTs on GPUs}

In this work, we focus on efficient FFT implementations on GPUs. We choose to focus on GPUs for several reasons. 
First, GPUs are one of the widely used accelerators present today.
In fact, GPUs power seven of ten fastest supercomputers in the world~\cite{top500Nov22}. 
Second, starting with an accelerated baseline like that on a GPU allows us to assess the efficacy of new accelerator solutions like 
\revise{PIM}
beyond existing state-of-the-art solutions. 
Finally, emerging commercial 
\revise{PIM}
solutions are coupled with GPUs~\cite{mi100PIM} allowing us a baseline architecture to assess. As such, we focus on analyzing efficient implementations of FFT on GPUs. 
We discuss the implications of our work for FFTs on other processors in \secref{related}. 

A common feature of efficient implementations of FFT on GPUs~\cite{rocfft,cufft,msft_fft_paper,msft_fft_paper_sc}) is that they anchor on effective use of local scratchpad (local data share/LDS on AMD GPUs or shared memory on other GPUs). That is, computing an FFT of size $N$ such that the $N$ input elements fit in LDS comprises loading the data in LDS and performing subsequent butterfly computations by accessing data from LDS and not main memory. By minimizing data movement and accessing data out of high-bandwidth on-chip scratchpad memory, high efficiency can be attained.

\begin{figure}[t]
    \centering
    \includegraphics[width=\linewidth]{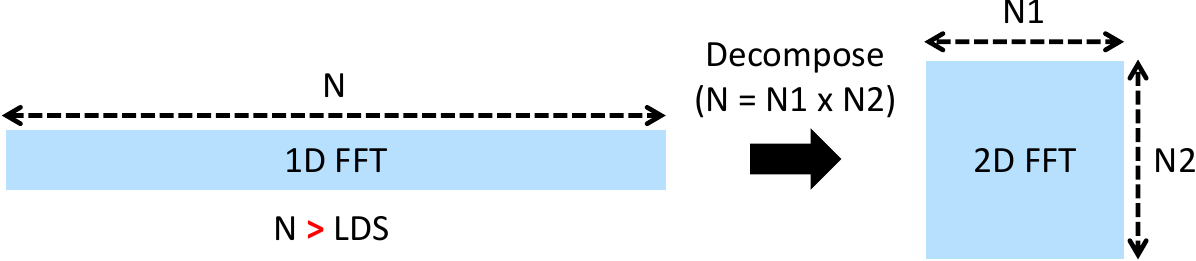}
    \caption{FFT decomposition.}
    \label{fig:bg_fft_decomposition_gpu}
    \vspace{-1.25\baselineskip}
\end{figure}

For FFT of size $N$ such that the $N$ input elements do not fit in LDS, existing FFT implementations decompose the problem into multi-dimensional (2D, 3D) space, processing each dimension sequentially. We depict a 2D decomposition in \figref{fig:bg_fft_decomposition_gpu}. The decomposition is guided by multiple factors, and we discuss some of the key factors. First, the decomposed components together form the original computation (that is, $N = N_1 \times N_2$). Second, the decomposed components are chosen to fit in LDS (that is, while $N$ elements do not fit in LDS, $N_1$ and $N_2$ elements, individually, fit in LDS). Note that, while a single GPU kernel is needed for FFT of size $N$ when $N$ input elements fit in LDS, for the depicted 2D decomposition, two GPU kernels are needed each representing a batched FFT computation. That is, $N_1$ (batch size) column FFTs of size $N_2$ followed by $N_2$ (batch size) row FFTs of size $N_1$. As such, decomposition leads to \textit{batched} FFT computations. Finally, FFT decomposition is used recursively whenever one of the dimensions does not fit in the LDS. 

\putssec{bckg_pim}{Commercial PIM Solutions}

Continued memory bandwidth demand, both from commercial and scientific workloads, has made it worthwhile for memory vendors to reassess, in a commercial context, processing-in-memory (PIM). 
PIM is a computing paradigm wherein portions of compute are offloaded to near-memory compute units to avail higher bandwidth (potentially an order of magnitude or more). As such, recently, multiple memory vendors have proposed commercial PIM designs more cognizant of industry constraints. In this work, we study a strawman PIM architecture which is an exemplar of recent commercial PIM designs~\cite{hynixPIM,samsungPIM}\footnote{While dissimilarities exist, these solutions have similar key architecture.}. As discussed above, as we study FFTs on GPUs, we focus on GPU-based PIM solutions which augment high bandwidth memory (HBM)~\cite{HBM-jedec}.

\begin{figure}[t]
    \centering
    \includegraphics[width=\linewidth]{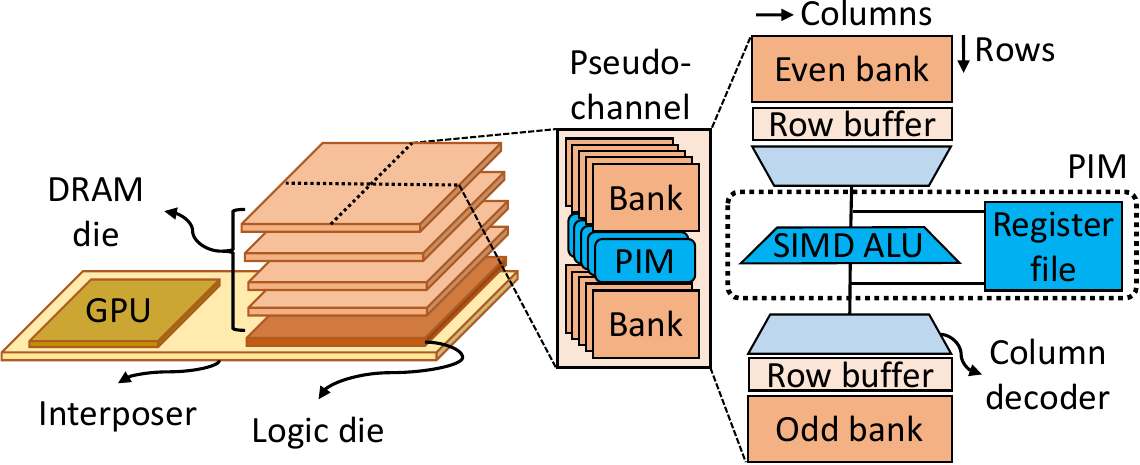}
    \caption{Strawman commercial HBM PIM architecture.}
    \label{fig:bg_hbm_dram_pim}
\end{figure}

Figure~\ref{fig:bg_hbm_dram_pim} depicts the strawman commercial PIM design we focus on in this work (henceforth referred to as PIM). To attain high bandwidth whilst incurring low energy/bit of data transfer, HBM is integrated with GPU within a single package, with communication amongst them via a silicon interposer. Each HBM module stacks multiples of four DRAM dies vertically, with a base logic die stacked below the memory dies. Each DRAM die comprises multiple pseudo channels, which are comprised of multiple banks. Banks within a pseudo channel share the data bus associated with the pseudo channel, while pairs of pseudo channels share a command bus. Read or write requests from the GPU typically access a single bank in a pseudo channel wherein, the address associated with the read/write request determines the  pseudo channel, bank, row, and column address within the bank to be accessed. A read (or write) request first causes the specified row within a bank to be activated (\textit{row activation}), which moves data in the row into a row-buffer structure associated with the bank. Next, data at specified column address is accessed within the row buffer via a \textit{column access} command. A row once activated (\textit{open row}) can process subsequent column accesses at lower overhead than accessing data in a separate row (necessitates another row activation). 

In order to offload computations to HBM, compute units (depicted as PIM) are shared between two banks in pseudo channel. Such sharing limits area overheads and potential memory capacity costs of adding compute to memory module. A PIM unit is composed of an ALU and a register file. The ALU width and register input/output is matched to the output width of the DRAM bank (e.g., 256 bits). In addition, the ALU is capable of operating on narrow words within single DRAM word (e.g., eight 32bit operands within a 256bit DRAM word). The register file serves as a scratchpad for computation in PIM units. PIM units are controlled via read/write like instructions from GPU such as add, subtract, multiply, etc. Software enforced data consistency (e.g., cache flush) is employed to ensure data dependency ordering between GPU and PIM instructions. 

As discussed, the key motivation for the above commercial interest in PIM designs is the memory bandwidth boost PIM avails. While a GPU read/write only accesses a single bank in a pseudo channel at a time given the shared data bus, by computing on data without traversing the data bus, multiple banks can be configured to compute on data at the same time (via multi-bank broadcast commands). This makes it possible to have a potential memory bandwidth multiplier of \texttt{\#}banks/2 with PIM over GPU. 
Emerging commercial PIM designs, however, issue PIM operations at half the rate of regular reads/writes to accommodate multi-bank broadcast commands~\cite{samsungPIM}.
As such, this bandwidth multiplier is about \texttt{\#}banks/4 in practice. For HBM memory with 16 banks per pseudo channel, this bandwidth boost is about 4x.
We discuss the bandwidth multiplier range in more detail in \ssecref{motiv_pim_mem_bw}. 
In addition to bandwidth amplification, 
PIM also avails considerable energy savings by not moving data (more than 50\% ~\cite{mi100PIM}).
\putsec{motiv}{Case for PIM Acceleration of FFT}

In this section, we motivate the consideration of PIM as a potential acceleration solution for FFTs. 
Our motivation anchors on two key observations which we expound on below. 
\revise{First, efficient FFT implementations on GPUs are memory bandwidth bound. Second, commercial PIM solutions avail memory bandwidth boost beyond that available at the GPU.}
Together these observations make a case for PIM to be a candidate accelerator for FFT. 

\putssec{motiv_fft_memory_bound}{FFT is Memory Bandwidth-bound}

\begin{figure}[t]
    \centering
    \includegraphics[width=\linewidth]{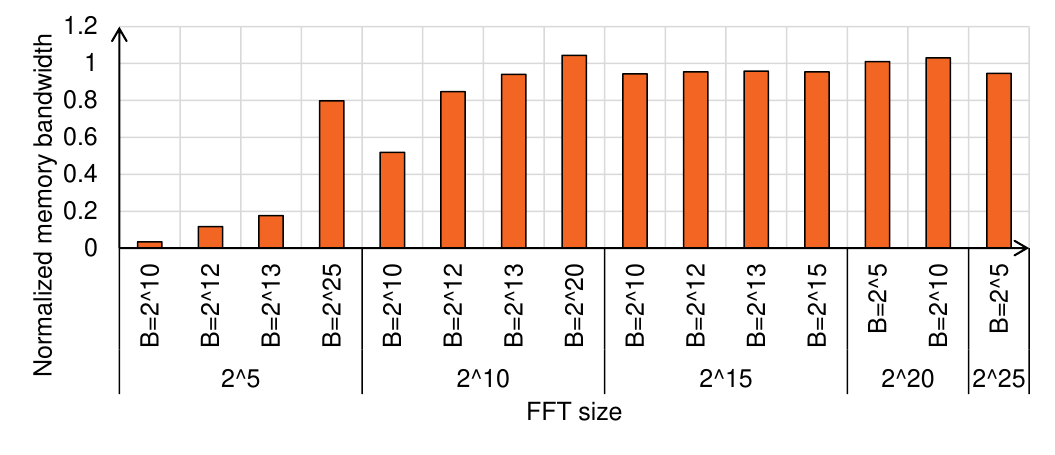}
    \caption{Efficient FFTs are memory bandwidth-bound.}
    \label{fig:motiv_fft_memory_bandwidth_bound}
    \vspace{-\baselineskip}
\end{figure}

\revise{To showcase the memory bandwidth boundedness of the efficient FFT implementations, we measure the memory bandwidth consumption of variety of FFT sizes and batch sizes (see \sssecref{proposal_base_eval_model} for setup details) and normalize to memory bandwidth consumption of the copy kernel from the BabelStream benchmark~\cite{babelstream,babelstream_on_amd} as shown in \figref{fig:motiv_fft_memory_bandwidth_bound}.  BabelStream is a synthetic GPU benchmark based on the STREAM benchmark for CPUs~\cite{stream_cpu}, which measures the sustainable memory bandwidth consumption to and from GPU memory. When running this benchmark with suitably large arrays, the data is streamed directly from GPU memory, and therefore memory bandwidth consumption of BabelStream
is a strong anchor to compare memory bandwidth boundedness of other computations.
} 
We observe two important trends in the data. First, as FFT size increases, memory bandwidth utilization increases and is high regardless of batch size 
(0.94$\times$ and 1.04$\times$ that of BabelStream for FFT size of $2^{10}$ with batch size $2^{13}$ and $2^{20}$, respectively). 
This is expected for as the FFT size increases, on-chip scratchpad/caches are not enough to hold inputs and further, the likelihood of decomposition increases leading to inputs being read/written multiple times. Second, even for smaller FFTs, as the batch size increases, the memory bandwidth utilization increases 
(up to 80\% that of BabelStream for FFT size of $2^5$ with batch size of $2^{25}$).
Larger batch sizes are, in effect, similar to large FFT sizes and hence manifest similar behavior. Overall, as the data shows, efficient GPU implementations considerably push the memory bandwidth and as such can benefit from memory bandwidth boost. 

\putssec{motiv_pim_mem_bw}{PIM Avails Memory Bandwidth Boost}

\begin{figure}[t]
    \centering
    \includegraphics[width=\linewidth]{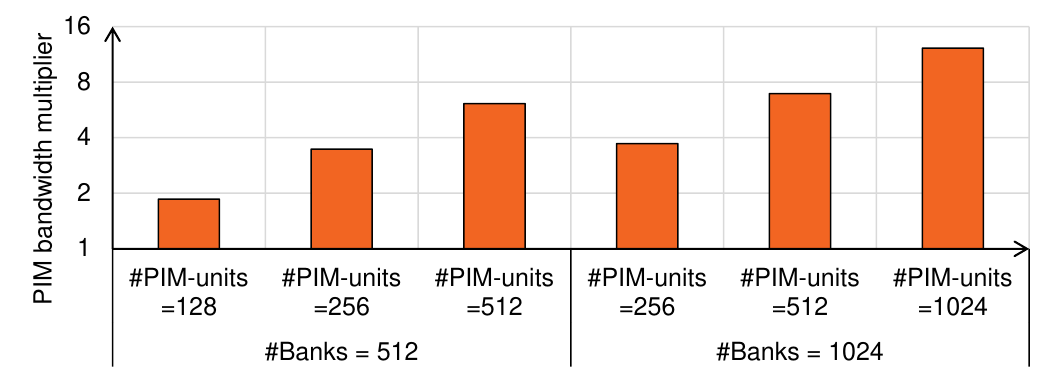}
    \caption{PIM bandwidth boost with GPU (optimistically) at 100\% bandwidth utilization.}
    \label{fig:motiv_pim_memory_bandwidth_mult}
    \vspace{-\baselineskip}
\end{figure}

In \figref{fig:motiv_pim_memory_bandwidth_mult}, we depict the memory bandwidth boost availed by 
emerging commercial PIM designs for forward-looking HBM memory
for selected representative FFT sizes and batch sizes. 
Given the current available PIM designs showcased on HBM2 memory, we assume an upcoming HBM3 memory~\cite{HBM3-jedec}
in our study for both PIM and baseline GPU (see \sssecref{proposal_base_eval_model}) as beside being forward-looking, it gives the best available bandwidth for GPU and presents a strong baseline for our work to improve upon. 
We consider different configurations such as baseline \#banks (512) along with hypothetical exploration of large \#banks (1024) due to increase in channels/stack or banks/channel. 
Additionally, we also vary \#PIM units provisioned. Note that, when \#PIM units are lower than \#banks, a PIM unit is shared amongst the banks. Overall, we observe that PIM can avail considerable memory bandwidth boost over GPU (up to 12x) by having multiple banks in a channel compute on data at the same time vs. GPU accessing a single bank at a time. 
Further, 
more banks/more PIM units favor PIM (higher bandwidth available by PIM units). 
Overall, this memory bandwidth boost can be 
beneficial for memory bandwidth bound workloads like FFT. 

\putssec{fft_pim_amenab}{PIM as Candidate Accelerator for FFT}
Given FFT is memory bandwidth bound (\ssecref{motiv_fft_memory_bound}) and PIM avails memory bandwidth boost (\ssecref{motiv_pim_mem_bw}), we believe that PIM is worthy of consideration as an accelerator for FFTs. To this end, we perform a detailed study of orchestrating FFTs on a strawman PIM architecture which is an exemplar of recent commercial PIM designs. We identify data mapping/orchestration considerations, challenges, and potential solutions to realize FFT acceleration with PIM. 
\putsec{proposal_base}{Baseline PIM-FFT}

In this section we discuss how we map the FFT computation to PIM. We first begin with an overview of key considerations which ought to be addressed when any computation is to be offloaded to PIM. Subsequently, we discuss how we address these considerations specifically for FFT by discussing data mapping and compute orchestration for offloading FFT to PIM. We term the resultant FFT PIM routine we discuss in this section as \base. 

\putssec{proposal_base_considerations}{PIM Offload Considerations}
As with any new accelerator, harnessing the strengths of the accelerator while avoiding stressing its shortcomings is necessary to harness performance. To that end, we first discuss the key considerations for offloading computation to PIM. 

\begin{figure*}[t]
    \centering
    \includegraphics[width=\textwidth]{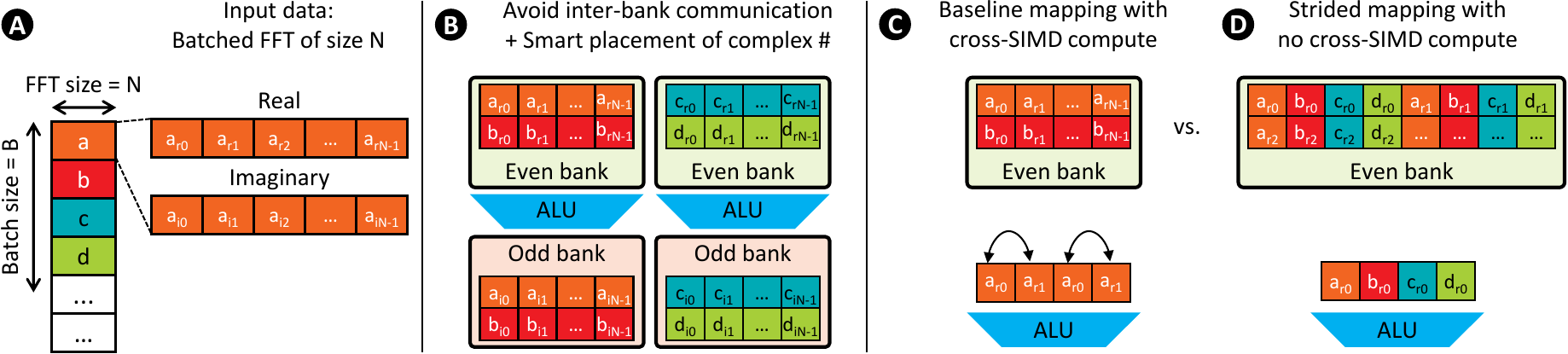}
    \caption{Proposed FFT data mapping for PIM.}
    \label{fig:base_pimacolaba_mapping_strided_packing}
    \vspace{-\baselineskip}
\end{figure*}

\noindent\textbf{Data Mapping.} As discussed in \ssecref{bckg_pim}, PIM provisions in-memory compute units. Specifically, our strawman PIM design, based on commercial PIM design~\cite{samsungPIM}, places a compute unit (e.g., ALU, register files) per two DRAM banks. Consequently, any interacting operands in the computation have to be placed in memory such that they are mapped to the same DRAM bank (or banks sharing a PIM compute unit). Further, PIM avails memory bandwidth boost by broadcasting the same command to multiple banks in the same pseudo-channel. If input/output data is interleaved appropriately across DRAM banks/channels, this broadcast feature can be harnessed. Note that avoiding inter-bank communication while harnessing PIM broadcast feature is an interesting balance. Finally, commercial PIM designs place a SIMD ALU near DRAM banks. While this aids in harnessing data parallelism, any cross-lane computations require shift operations 
which can be costly in DRAM technology due to the limited number of metal layers.
These considerations have to be addressed while offloading a computation to PIM to harness acceleration. Note, even efficient computations on GPUs have data mapping considerations (e.g., reading/writing memory contiguously) to extract optimal performance. 

\noindent\textbf{Compute Orchestration.} PIM computations are kicked off by launching \textit{pim kernels}, which are like existing GPU kernels except they issue \textit{pim instructions}. A \textit{pim instruction} has the effect of enqueuing a \textit{pim command} at the memory controller which in turn instructs PIM unit to execute either computation (e.g., add, multiply, etc.) or data movement (read from row-buffer to register, write from register to row-buffer, etc.) along with necessary row activation. As discussed in \ssecref{bckg_pim}, each PIM compute operation is a SIMD operation (e.g., eight 32bit operations over 256bit DRAM word). Finally, since memory channels are independently controlled in GPUs with multiple memory controllers, different groups of threads (e.g., workgroups or thread blocks) issue commands to different channels (commands broadcasted to banks within a channel). 

Overall, programmer first decides data mapping for a computation to maximize harnessing of PIM strengths (e.g., command broadcasts) and avoid stressing PIM shortcomings (e.g., cross SIMD compute, inter-bank communication). Subsequently, a \textit{pim kernel} which expresses the computation orchestration on in-memory compute units is launched. 

\putssec{proposal_base_data_mapping}{PIM FFT Data Mapping}
We discuss in this section the data mapping considerations specific to FFT
and depict our choices with the help of \figref{fig:base_pimacolaba_mapping_strided_packing}. Note that in our work we focus on complex DFT and as such, input is comprised of complex numbers with both real and imaginary components \circled{A}. Consequently, data mapping consideration for FFT involves deciding how the $N$ complex numbers which are the inputs to the FFT computation are placed/mapped in memory. We discuss our design decisions given the data mapping constraints we discussed in \ssecref{proposal_base_considerations}.

\putsssec{base_avoid_interbank}{Avoiding Inter-bank Communication}
As discussed in \ssecref{bckg_fft}, for a FFT of size $N$, elements interact with each other at different strides in different steps. This makes mapping of these elements while avoiding inter-bank communication challenging. To tackle this, we consciously choose in our \base design to consider FFT sizes such that number of elements $N$ fit in a pair of DRAM banks that share a PIM unit. This avoids any inter-bank communication as all interacting elements are mapped to banks with a shared ALU.  This limits the maximum FFT size that we can tackle in our \base design to $2^{21}$ with single-precision elements (we will overcome PIM FFT size reach with alternate strategies in subsequent sections).  Furthermore, we harness the fact that banks share ALU, to opportunistically place real and imaginary components of a given element in even and odd DRAM banks, respectively \circled{B}. This allows us to access both components at the same time in our computations without incurring costly row-opens.

\putsssec{base_avoid_cross_simd}{Avoiding Cross-SIMD Compute}
Inter-element interaction in FFT computation can also lead to cross-SIMD computation which is costly in PIM (baseline mapping) \circled{C}. To avoid these, we 
choose in our \base design to pack $N$ elements belonging to a single FFT in single SIMD lane (termed strided mapping) \circled{D}. Note, this reduces the maximum FFT size that we can tackle in our \base design further to $2^{18}$ (driven by SIMD width and DRAM row buffer size). Furthermore, this can also lead to memory wastage if all SIMD lanes are not utilized. We discuss how we tackle this next. 

\putsssec{base_avoid_waste}{Harnessing PIM Broadcasts and Avoiding Memory Wastage}
As discussed in \ssecref{bckg_efficient_gpu}, FFTs decomposition leads to batched FFT computations. We employ such batching to both avoid memory wastage due to our data mapping design choices so far and harness PIM broadcast feature/memory bandwidth boost. That is, first, while we pack a single FFT in one SIMD lane, batching avails us of concurrent FFTs which can occupy the residual SIMD lanes and avoid memory wastage. Second, batching also allows us to spread available FFT batches across channels and banks, thus allowing us to broadcast same command across channels/banks and avail memory bandwidth boost of PIM. That is, we can compute multiple FFTs in different banks/channels concurrently by broadcasting the same PIM instructions/commands (as we discuss next). 

\begin{figure}[t]
    \centering
    \includegraphics[width=\linewidth]{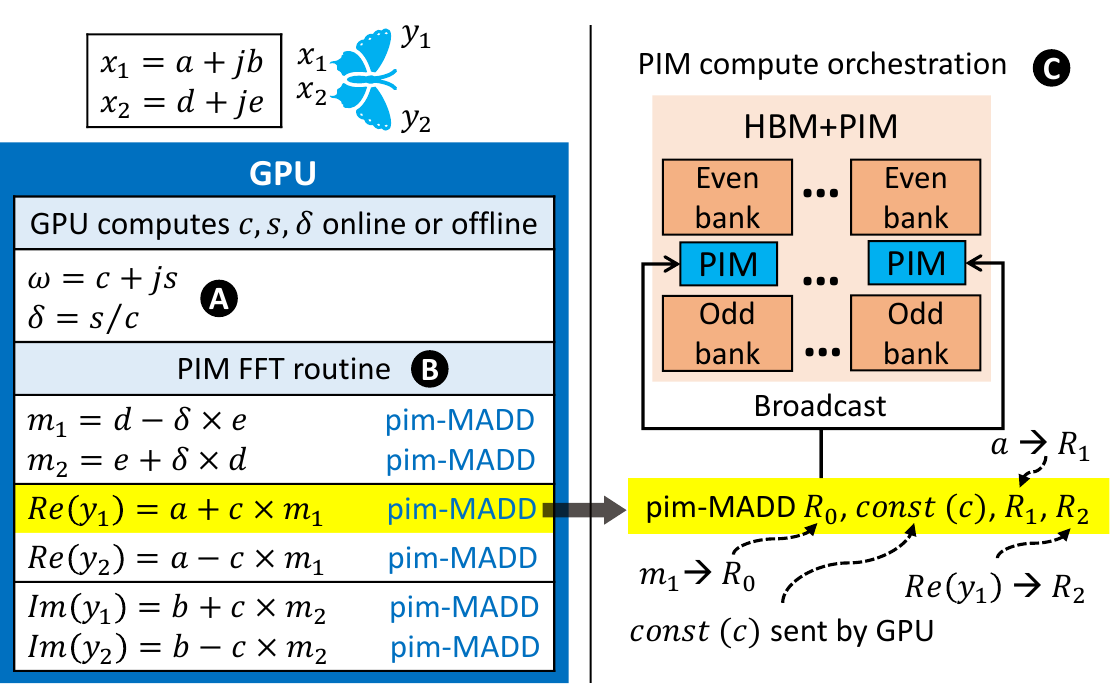}
    \caption{PIM FFT routine \& orchestration.}
    \label{fig:model_pim_gpu_interaction}
    \vspace{-\baselineskip}
\end{figure}

\putssec{proposal_base_orchestration}{PIM FFT Routine}
As discussed in \ssecref{bckg_fft}, the key building block of FFT is the butterfly computation. We depict in \figref{fig:model_pim_gpu_interaction} orchestration of a single butterfly computation in our \base FFT routine. As also discussed in \ssecref{bckg_fft}, the butterfly computation comprises complex number multiplication and addition. Further, we depict how the required computation can be factored into six \textit{pim-MADD} commands (multiply and add) \circled{B} along with an online or offline computation of twiddle factor components \circled{A}. As discussed in \sssecref{base_avoid_waste}, using our \textit{pim-kernel}, we broadcast these commands to banks in a channel (and to multiple channels) to concurrently compute multiple FFTs in a batch \circled{C}. 

\putssec{proposal_base_eval}{Performance Analysis of \base}
In this section, we evaluate the performance of our proposed PIM FFT mapping, \base and further dive deep into how the design decisions we have discussed so far play out. We first start with motivating and discussing our performance models. Next, we discuss the effects of our data mapping choices and end with speedup analysis of our proposed \base routine vis-a-vis GPU. 

\putsssec{proposal_base_eval_model}{Setup and Performance Model}
We outline in this section our system setup (used for \figref{fig:motiv_fft_memory_bandwidth_bound} and more) and performance models for FFT computations on GPU and PIM. In our work, we choose to analyze performance using analytical models for several reasons. First, PIM is only yet available as part of functional prototypes (e.g., 
Lee et al.~\cite{samsungPIM} couple 
a modified HBM2 with PIM capability with a off-the-shelf GPU as a functional prototype). Second, we aim to study performant, highly efficient FFT GPU implementations for a variety of sizes and whilst considering software optimizations (see below). 
This makes relying on GPU simulators difficult. 
Finally, as we discuss below, our setup provisions a much stronger baseline to show PIM benefits over.  

\noindent\textbf{System Setup.}
Our system setup consists of an AMD Instinct\textsuperscript{\texttrademark} MI210 Accelerator comprising GPU with 104 compute units (or GPU cores) and four stacks of HBM2E memory for a total capacity of 64GB and a peak memory bandwidth of 1638.4 GB/s~\cite{mi210}. We profile/study behavior of FFT on GPU using rocFFT~\cite{rocfft} library which is part of AMD's software ecosystem based on ROCm. We use Omniperf~\cite{omniperf}, a system performance profiling tool for machine learning/HPC workloads running on AMD MI GPUs, to study behavior of FFT kernels and gather performance statistics (e.g., reads/writes to HBM memory, memory bandwidth, etc.). 
Finally, we run the copy kernel from BabelStream using the maximum problem size to measure the memory bandwidth consumption as discussed in \ssecref{motiv_fft_memory_bound}.

\noindent\textbf{GPU Performance Model.} Given the memory bandwidth boundedness of FFT computations (\ssecref{motiv_fft_memory_bound}), for our GPU performance model, we assume that the GPU execution time is only limited by available memory bandwidth (i.e., we assume the compute to be free). Further, we observe that as FFT is decomposed, transpose kernels can be used in certain situations to improve the access patterns for computations. Increasingly as such kernels can be fused with FFT computation kernels~\cite{msft_fft_paper,msft_fft_paper_sc}, we subtract out the effects of transpose kernels to assume an even stronger GPU baseline.  

\begin{figure}[t]
    \centering
    \includegraphics[width=\linewidth]{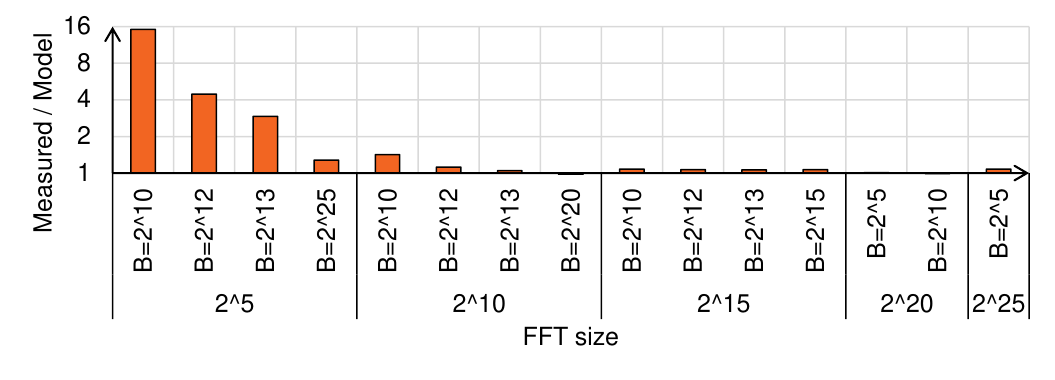}
    \caption{Fidelity of our GPU performance model.}
    \label{fig:model_host_fidelity}
\end{figure}

\figref{fig:model_host_fidelity} shows the fidelity of our GPU performance model for a variety of representative FFT sizes and batch sizes (same sizes shown in \figref{fig:motiv_fft_memory_bandwidth_bound}). For our GPU performance model, we measure memory reads and writes only for FFT compute kernels (no transpose kernels) and 
\revise{assume the maximum memory bandwidth utilization reported by the copy kernel from BabelStream for the baseline GPU (\ssecref{motiv_fft_memory_bound}).}
We compare this execution time to actual measured runtime. \figref{fig:model_host_fidelity} depicts that as FFT size or batch size increases, the computation gets increasing memory bandwidth bound and our performance model tracks well with measured execution time. 
For FFTs with small sizes or small batch sizes, where the computation is not memory bandwidth bound, our model projects a far more optimistic execution time than possible. We still use our proposed performance model to assess PIM speedups across all sizes to keep a unified model. Note, this means that, given our detailed performance model for PIM (see below), PIM benefits for small FFT sizes/batch sizes, will likely be higher than what we discuss below. 
Finally, as shown 
in \ssecref{proposal_colab_eval}, the combination of small FFT sizes with small batch sizes is uncommon in our proposed techniques.

\noindent\textbf{PIM Performance Model.} We take a detailed DRAM command orchestration approach for our PIM performance model. That is, for a given FFT size, we deduce data mapping (\ssecref{proposal_base_data_mapping}) and orchestration (\ssecref{proposal_base_orchestration}) necessary. Subsequently, we deduce the exact DRAM commands needed to orchestrate the computation (e.g., including row activations). We assume the parameters listed in \tabref{tab:model_params} for our model. Note that we assume a PIM-aware GPU which can issue \textit{pim-instructions} and \textit{pim-commands} at issue-rate. With the available thread parallelism at the GPU, we believe this to be a reasonable assumption. 

\begin{table}[t]
\centering
\caption{Parameters for performance model.}
\label{tab:model_params}
\resizebox{\columnwidth}{!}{%
\begin{tabular}{|l|l|}
\hline
\textbf{\#Banks per Stack (4-high)}      & 512~\cite{HBM3-jedec}                                 \\ \hline
\textbf{Bandwidth per Pin}               & 4.8 Gb/s~\cite{HBM3-jedec}                            \\ \hline
\textbf{GPU Memory Bandwidth per Stack} & 614.4 GB/s~\cite{HBM3-jedec}                          \\ \hline
\textbf{Row Buffer Size}                 & 1024 B~\cite{HBM3-jedec}                              \\ \hline
\textbf{DRAM Parameters}                 & tRP = 15ns, tCCDL=3.33ns, tRAS=33ns~\cite{HBM3-jedec} \\ \hline
\textbf{PIM Parameters} & \begin{tabular}[c]{@{}l@{}}\#PIM Units per Stack = 256\\ \#PIM Registers per ALU = 16\end{tabular} \\ \hline
\end{tabular}%
}
\end{table}

\putsssec{proposal_base_eval_data_mapping}{Data Mapping Evaluation}

\begin{figure}[t]
    \centering
    \includegraphics[width=\linewidth]{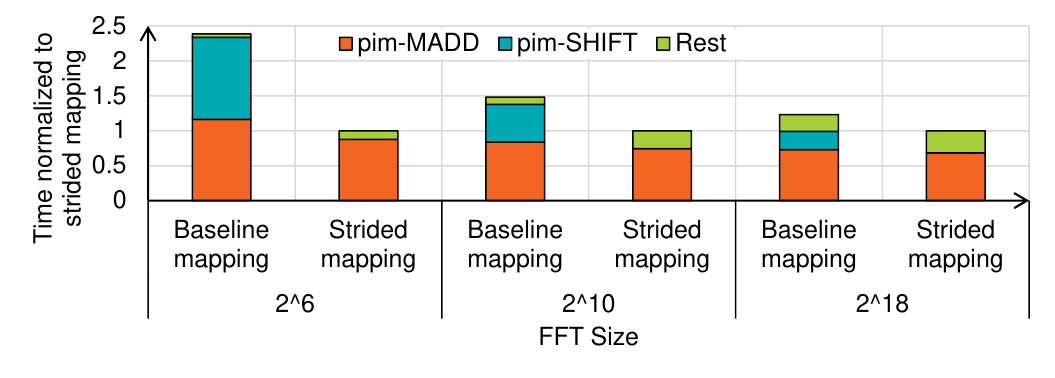}
    \caption{Strided mapping vs. baseline mapping.}
    \label{fig:base_pimacolaba_strided_packing_perf}
\end{figure}

While in \ssecref{proposal_base_data_mapping} we discussed various data mapping considerations, 
in this section we evaluate baseline mapping and strided mapping.
\figref{fig:base_pimacolaba_strided_packing_perf} depicts evaluation of our proposed strided data mapping (\sssecref{base_avoid_cross_simd}) to baseline data mapping. We depict along the y-axis the execution time normalized to strided mapping for each FFT size. Further, we break down the execution time into that spent executing two key PIM instructions (\textit{pim-MADD} and \textit{pim-SHIFT}) and rest of the time as \textit{Rest} (contains DRAM row-open overhead, data movement from register to row-buffer, etc.). Note that only baseline mapping scheme relies on costly shift instructions. 
As \figref{fig:base_pimacolaba_strided_packing_perf} depicts, strided mapping, by avoiding \textit{pim-SHIFT} instructions, is superior to baseline mapping. As FFT size increases, we do see these schemes get closer in performance albeit strided mapping still maintains an edge. This is so, as FFT size increases, portion of computation needing cross-lane interaction and hence \textit{pim-SHIFT} drops. While this is so, note also that functionality like lane shifts is hard to provision in DRAM technology. Overall, our proposed strided mapping eliminates the need for costly \textit{pim-SHIFT} commands which translates to significant reduction in FFT execution time on PIM, especially for the small FFT sizes.

\begin{figure}[t]
    \centering
    \includegraphics[width=\linewidth]{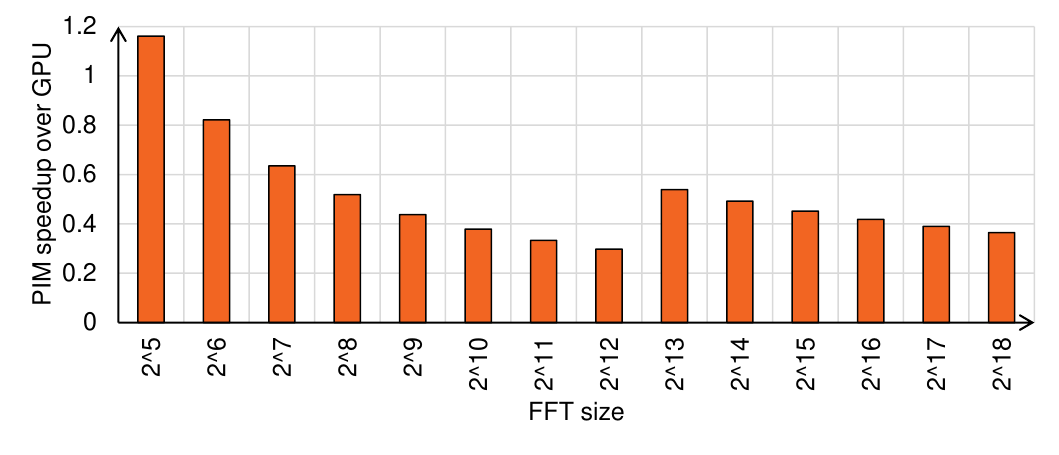}
    \caption{PIM speedup under \base.}
    \label{fig:base_pimacolaba_pim_speedup}
    \vspace{-\baselineskip}
\end{figure}

\putsssec{proposal_base_eval_speedup}{Speedup with \base}
Finally, we evaluate the performance of our \base FFT routine given the data mapping and orchestration decisions we have made thus far. We depict speedup of \base over GPU for various FFT sizes in \figref{fig:base_pimacolaba_pim_speedup} up to the max size we can support in PIM ($2^{18}$). As the figure depicts, despite careful data mapping and orchestration, except for small sizes ($2^5$), \base FFT routine incurs considerable slowdown vis-a-vis GPU (average slowdown of 52\%). 
We believe the primary reason for this is that while FFT is memory bandwidth bound on GPU, FFT manifests compute-boundedness in PIM. This is so as when FFT is mapped to GPU, only reads/writes consume memory bandwidth. In contrast, when FFT is mapped to PIM, every operation in FFT computation is now a PIM compute command (e.g., \textit{pim-MADD}). Further, PIM compute throughput is typically lower than GPU. As an example, for our MI210 GPU with four HBM2e memory stacks, the peak single-precision PIM throughput is about seven times lower as compared to the GPU which considerably stresses PIM compute\footnote{As GPU compute throughput typically increases faster than memory bandwidth, this performance gap will exist or can potentially get wider in the future.}. Overall, this analysis points to a more nuanced approach to harness PIM for FFT than a binary decision to use or not use PIM for the entire computation.

\putsec{proposal_colab}{Collaborative PIM-FFT}

We discussed in \secref{proposal_base} how even with careful data mapping and orchestration, our PIM FFT routine so far (\base) incurs considerable slowdown vis-a-vis GPU. To tackle this, in this section, we motivate, propose, and analyze an alternate strategy to offload FFT computations to PIM which harnesses \base but moves away from a binary offload decision (all or none of computation offloaded to PIM) to a more judicious offload mechanism where GPU and PIM collaborate to complete a FFT computation. We term the resultant FFT PIM mapping as \colab.

\putssec{proposal_colab_motiv}{Collaborative Decomposition}

Our proposed \colab PIM mapping is influenced by a confluence of several observations. First, our analysis in \secref{proposal_base} showed that there are some FFT sizes ($2^5$) where \base does provide performance benefit. Further, in other cases ($2^6$), while \base is slower by a small amount, by offloading to PIM, we can harness data movement savings at a small performance cost. That is, when GPU performs the computation, data is read/written to HBM, and resultant energy expenditure is incurred. Instead, if the computation were to be offloaded to PIM, the said data movement and resultant energy can be saved at a small performance cost. As such, we can attain performance acceleration/data movement savings if we have an avenue to invoke PIM for certain sizes only. Second, existing efficient FFT implementations already provide our desired mechanism. That is, as discussed in  \ssecref{bckg_efficient_gpu} efficient FFT implementations already decompose a problem into constituent components to better harness GPU scratchpad size. We propose augmenting this existing decomposition mechanism to invoke both GPU (existing kernels) and PIM component (\base). We term this strategy \textit{collaborative decomposition}.

\begin{figure}[t]
    \centering
    \includegraphics[width=\linewidth]{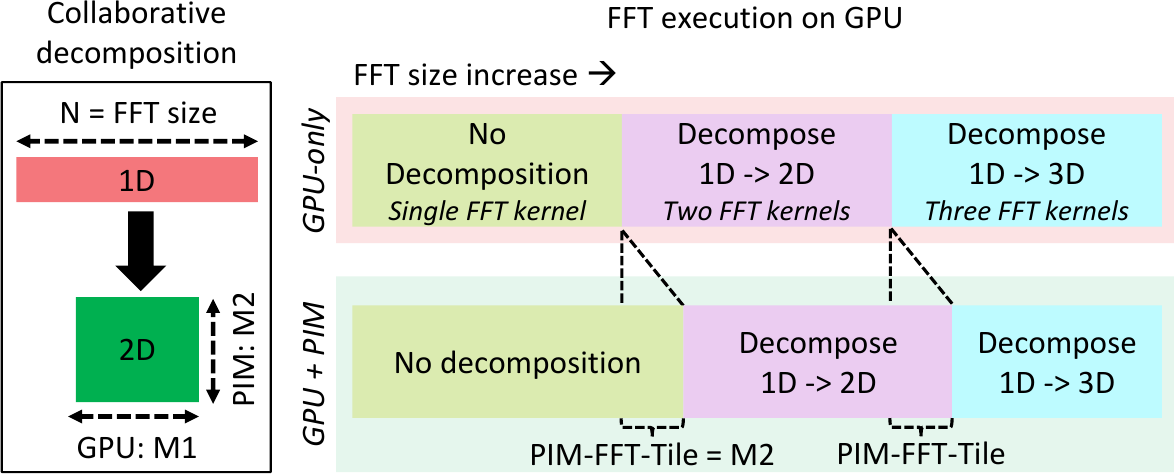}
    \caption{Collaborative decomposition in \colab.}
    \label{fig:collaborative_pimacolaba_decomposition_and_host_benefit}
    \vspace{-\baselineskip}
\end{figure}

\figref{fig:collaborative_pimacolaba_decomposition_and_host_benefit} depicts collaborative decomposition (left). As depicted, a given FFT of size $N$ is decomposed into GPU kernel (FFT of size $M1$, batch $M2$) and PIM kernel (FFT of size $M2$, termed \textit{PIM-FFT-Tile}, batch $M1$). Our choice of \textit{PIM-FFT-Tile} is driven by a simple algorithm: we pick the PIM FFT offload size such that we end up with same or less total \#kernels invoked (PIM or GPU). In the presence of multiple choices, we pick the most efficient \textit{PIM-FFT-Tile} (note, this can be analyzed once, offline). 

\figref{fig:collaborative_pimacolaba_decomposition_and_host_benefit} also depicts FFT kernels invoked by GPU as FFT size increases from left-to-right for baseline GPU and for \colab. For baseline GPU, available scratchpad space (LDS) dictates \#kernels to be invoked and as such, as we go from left-to-right, GPU invokes increasingly more kernels (one, two, three). With collaborative execution, the net effect is we shift the (size-range)-to-(kernel-count) association boundaries and effectively shrink the region where three GPU kernels are needed as depicted in \figref{fig:collaborative_pimacolaba_decomposition_and_host_benefit}. 
Overall, our proposed collaborative decomposition strategy has several benefits. First, it employs PIM judiciously and as we show in \ssecref{proposal_colab_eval} stands to better harness PIM acceleration. Second, it does so while piggybacking on existing efficient FFT mechanism of FFT decomposition. Finally, it avails considerable data movement savings.

\putssec{proposal_colab_eval}{Performance Analysis of \colab}

\putsssec{proposal_colab_speedup}{Speedup of \colab}

\begin{figure}[t]
    \centering
    \includegraphics[width=\linewidth]{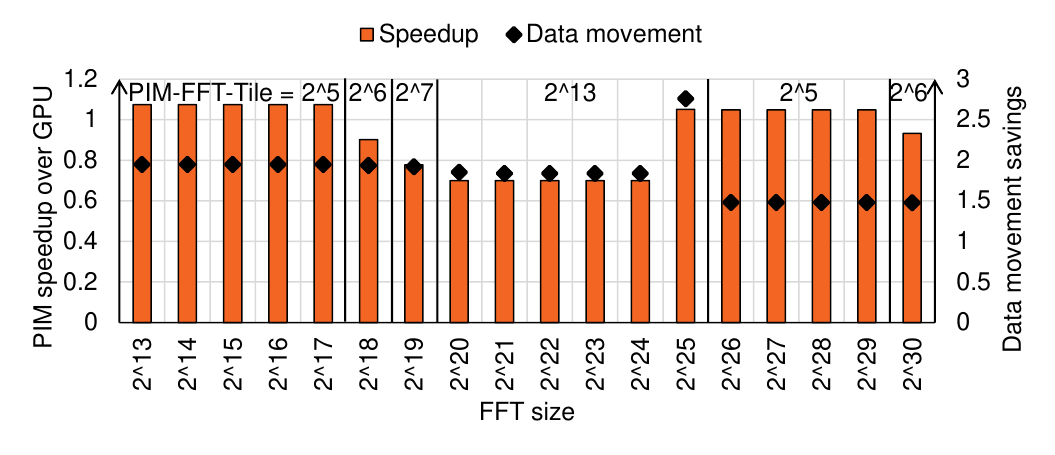}
    \caption{PIM speedup and data movement savings for \colab and \textit{PIM-FFT-Tile} used.}
    \label{fig:collaborative_pimacolaba_pim_speedup}
    \vspace{-\baselineskip}
\end{figure}

We evaluate the performance of \colab and depict resultant speedup over GPU for various FFT sizes in \figref{fig:collaborative_pimacolaba_pim_speedup}. Notice that the size range we depict here is different from \base speedup results (\figref{fig:base_pimacolaba_pim_speedup}). This is so, as first, the key tenet for \colab is to invoke PIM judiciously. As such, for FFT sizes, where GPU invokes single kernel (less than $2^{13}$ on our setup) and is already efficient, \colab does not harness PIM. As such, the performance is the same as baseline GPU. Second, \colab harnesses both PIM and GPU and as such, the max size we can harness PIM for increases to max FFT size the GPU memory can support ($2^{30}$ for our setup).  Overall, as the figure depicts, by judiciously using PIM, \colab does considerably better than \base.  For several sizes, by harnessing PIM only where it makes sense, we attain speedup over GPU. In cases where we do not attain performance, \colab presents a trade-off: data movement savings of \revise{up to 2.67$\times$} at some performance cost. 

\putsssec{proposal_colab_analysis}{Compute behavior of \colab}

\begin{figure}[t]
    \centering
    \includegraphics[width=\linewidth]{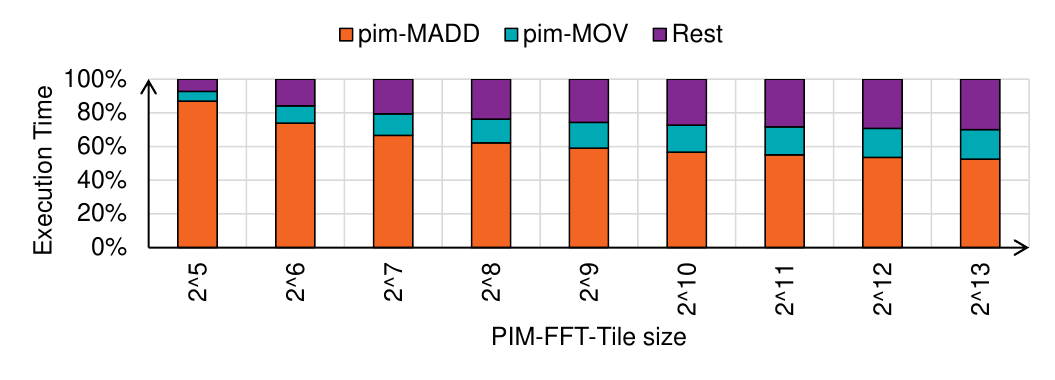}
    \caption{\colab is dominated with PIM compute. 
    }
    \label{fig:motiv_pim_compute_ops_bottleneck}
    \vspace{-\baselineskip}
\end{figure}

While \colab considerably improves over \base, we analyze \colab further to understand its behavior and deduce additional optimizations. \figref{fig:motiv_pim_compute_ops_bottleneck} depicts \colab execution time proportioning for FFT sizes we employ as \textit{PIM-FFT-Tiles}. We break down the execution time into that spent executing PIM computation (\textit{pim-MADD}), data movement local to PIM unit (\textit{pim-MOV}, move data from register to row-buffer and vice versa) and rest of the time as \textit{Rest} (contains DRAM row-open overhead etc.). 
We observe herein that the majority of PIM execution time is spent on compute operations or \textit{pim-MADD} commands, the building blocks of the butterfly computation. Specifically, \textit{pim-MADD} commands represent an average of 76\% of the PIM compute commands and an average of 54\% of the total PIM execution time \revise{(average not shown for space reasons)}. 
The remaining time is spent moving the FFT data in and out of PIM registers. As such, any further improvements in PIM execution will have to lower resultant compute operations. As a limit study, if \base used one \textit{pim-MADD} command instead of six we use now (\ssecref{proposal_base_orchestration}), this can lead to a speedup of up to 4.22$\times$ (figure omitted for space reasons). To this end, we next focus on software and hardware optimizations which aim to lower PIM compute commands. 

\putsec{proposal_opt}{\PNAME}

We discussed in \sssecref{proposal_colab_analysis} how optimizing PIM compute commands is necessary to attain further PIM acceleration. To this end, we motivate and analyze both software optimization and hardware augmentation which directly aim to lower PIM compute commands. We also discuss how these can be combined to further lower PIM compute commands. We term the resultant FFT PIM mapping, which harnesses collaborative decomposition and our proposed optimizations as \PNAMEBOLD. 

\begin{figure}[t]
    \centering
    \includegraphics[width=\linewidth]{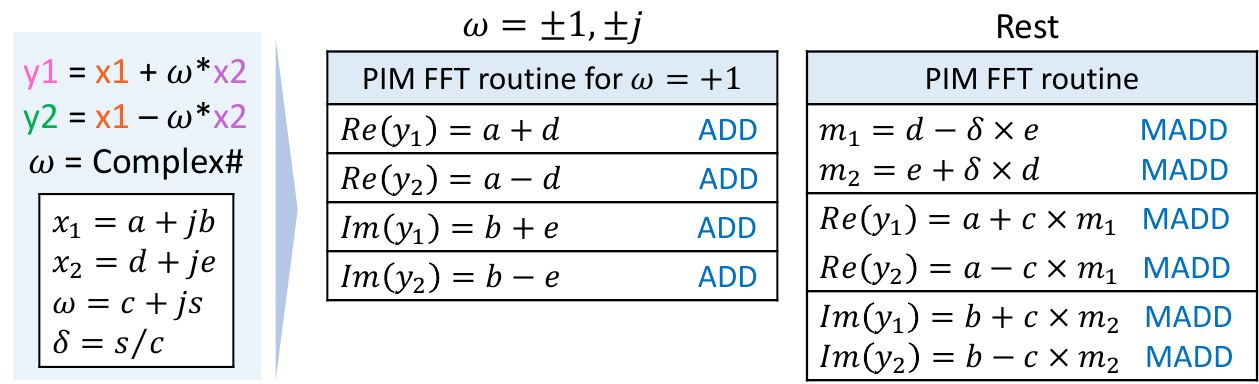}
    \caption{Twiddle factor aware PIM orchestration to reduce the number of PIM compute commands.}
    \label{fig:optmiz_sw}
\end{figure}

\putssec{proposal_opt_sw}{Twiddle Factor Aware PIM Orchestration}
We analyze the values of twiddle factors involved in FFT computation at different stages and notice that we can harness these differing values to lower PIM compute commands. Specifically, we notice that first, while twiddle factors needed is a function of FFT size $N$, they are deterministic based on FFT step being computed on. That is FFT size $N$ subsumes the twiddle factors needed for size $N-1$. Second, twiddle factors $1$ or $-j$ are used in initial FFT steps. For these specific values, as our \figref{fig:optmiz_sw} depicts, we can reduce the PIM compute commands needed for a single butterfly computation from six \textit{pim-MADD} to four \textit{pim-ADD operations}. As these values are used repeatedly, by enabling twiddle factor aware PIM orchestration from GPU, as we will show below, average number of \textit{pim-MADD} commands per butterfly can be lowered. 

\putssec{proposal_opt_hw}{PIM Augmentations for FFT}

\begin{figure}[t]
    \centering
    \includegraphics[width=\linewidth]{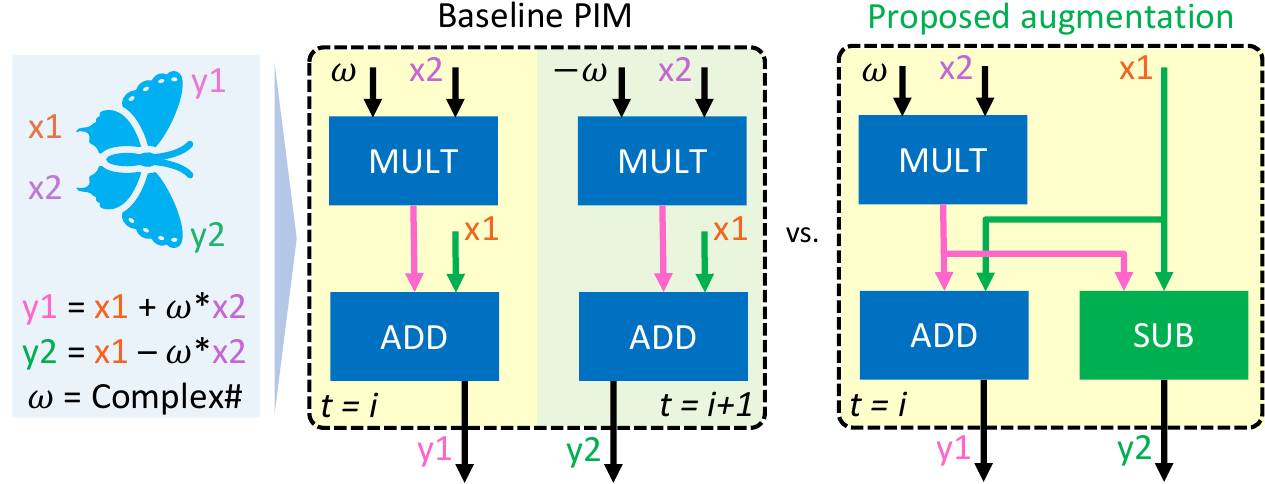}
    \caption{PIM ALU augmentations to reduce the number of PIM compute commands.}
    \label{fig:optmiz_hw_pim_augment}
    \vspace{-1.25\baselineskip}
\end{figure}

Analyzing the operations in butterfly computation we observe that the result of the same multiply operation ($\omega \times x_2$) is reused in an addition and a subtraction. Using baseline PIM ALU design, while we reuse the result of the multiplication, addition and subtraction take two PIM commands to be orchestrated by the GPU. Instead, we propose to augment the PIM ALU unit, as depicted in \figref{fig:optmiz_hw_pim_augment} such that a single PIM command realizes not just multiplication and addition (\textit{pim-MADD}) as the baseline PIM design supports, but also an additional subtraction. This augmentation has the net effect of bringing down the number of \textit{pim-MADD} commands per butterfly to four instead of six, independent of the used twiddle factor. Our proposed PIM command does not affect PIM orchestration; however, this necessitates additional write port to PIM ALU register file. 

\putssec{proposal_opt_both}{Combining Optimizations}
Note, our above PIM ALU augmentation helps us further enhance twiddle factor aware orchestration. First, in computations with twiddle factors of $1$ and $-j$, each butterfly can now be computed using two PIM commands. Furthermore, for butterflies with $\pm 1 / \sqrt{2}$ twiddle factor, the symmetry of the real and imaginary parts can be exploited to reduce the number of PIM commands to three. We study this combination in our analysis below. 

\putssec{proposal_opt_eval}{Performance Analysis of \PNAME}

\putsssec{proposal_opt_sw}{Optimized \textit{PIM-FFT-Tile}}

\begin{figure}[t]
    \centering
    \includegraphics[width=\linewidth]{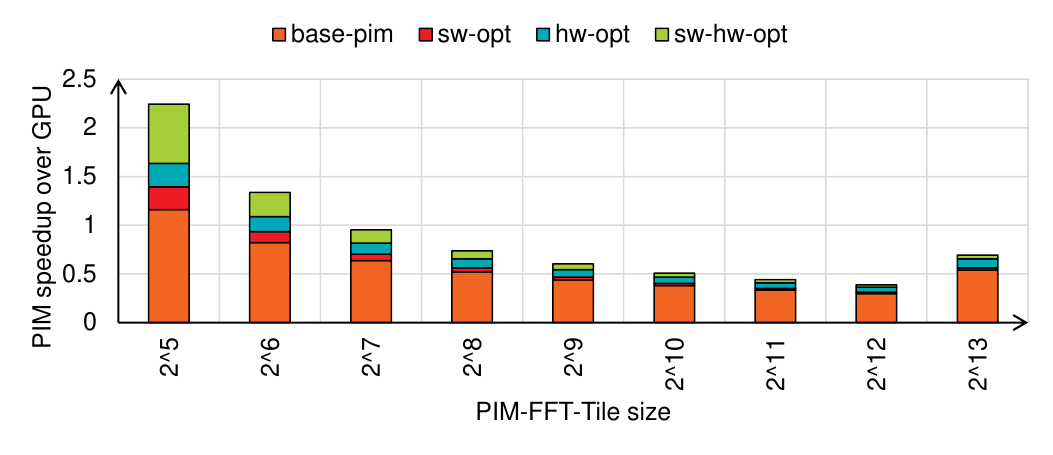}
    \caption{Optimized \textit{PIM-FFT-Tile}.}
    \label{fig:optmiz_pim_speedup}
\end{figure}

We depict in \figref{fig:optmiz_pim_speedup} the speedups attained by various optimizations we discuss above for FFT sizes we employ as \textit{PIM-FFT-Tiles}. We denote our twiddle factor aware orchestration as \textit{sw-opt}, PIM ALU augmentation as \textit{hw-opt}, combining the two as \textit{sw-hw-opt}. 
Note, we discuss the effects of this optimized \textit{PIM-FFT-Tiles} on the overall computation in the next section. 
We observe in \figref{fig:optmiz_pim_speedup} that \textit{sw-opt} improves the performance of small \textit{PIM-FFT-Tiles} ($\le2^6$) albeit with diminishing returns as \textit{PIM-FFT-Tile} increases. This is because as the FFT size increases the proportion of twiddle factors we optimize for drops. Overall, \textit{sw-opt} addresses the PIM compute bottleneck to some extent by reducing the average number of \textit{pim-MADD} commands to range from 4.85 and 5.54 per butterfly (vs. 6). Compared to \textit{sw-opt}, \textit{hw-opt} leads to better speedups. This is because \textit{hw-opt} benefits all butterfly computations regardless of twiddle factors. This enables \textit{hw-opt} to tackle the PIM compute bottleneck by reducing the average number of \textit{pim-MADD} commands to four per butterfly across all \textit{PIM-FFT-Tiles}. Finally, combining the two with our \textit{sw-hw-opt} gets us the best of both optimizations and leads to even lower  \textit{pim-MADD} commands per butterfly (2.67 to 3.46). Overall, compared to GPU, \textit{sw-hw-opt} provides higher acceleration for a range of \textit{PIM-FFT-Tiles} and therefore enables more options to be used when collaboratively decomposing a given FFT between GPU and PIM.

\putsssec{proposal_opt_sw}{Speedup of \PNAME}

\begin{figure}[t]
    \centering
    \includegraphics[width=\linewidth]{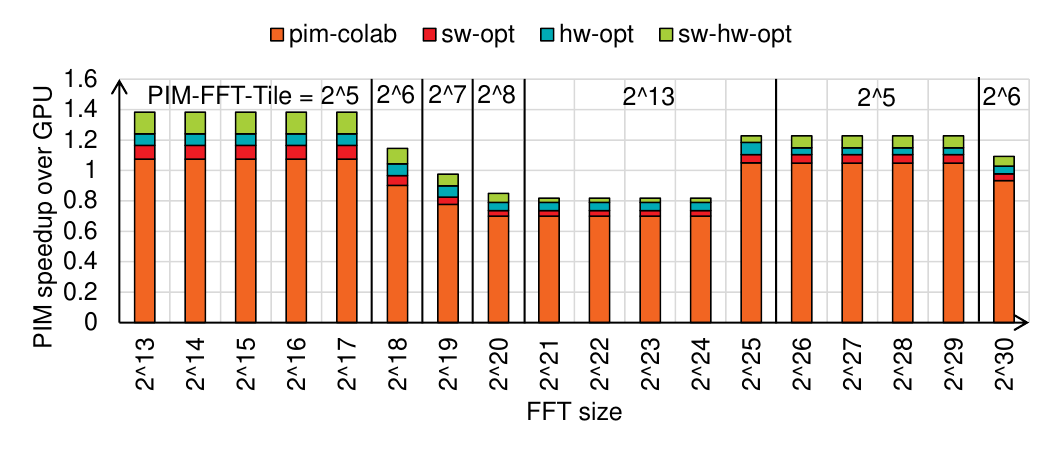}
    \caption{\PNAME speedup with optimized \textit{PIM-FFT-Tile}}
    \label{fig:overall_pim_speedup}
\end{figure}
Next, we analyze how overall FFT performance looks (combining our optimized \textit{PIM-FFT-Tile} with our collaborative decomposition approach). As discussed, we term this \PNAME. We depict this in \figref{fig:overall_pim_speedup}. Using our \textit{sw-opt} and \textit{hw-opt} approach we see speedups up to a max of 1.16$\times$ and  1.24$\times$, respectively. Combining the two, with \PNAME, we see a max speedup of 1.38$\times$. An interesting benefit of our optimizations is that they increase the possible \textit{PIM-FFT-Tile} options available as depicted in  \figref{fig:overall_pim_speedup}. 

\putssec{proposal_opt_data_mov}{Data Movement Savings of \PNAME}

\begin{figure}[t]
    \centering
    \includegraphics[width=\linewidth]{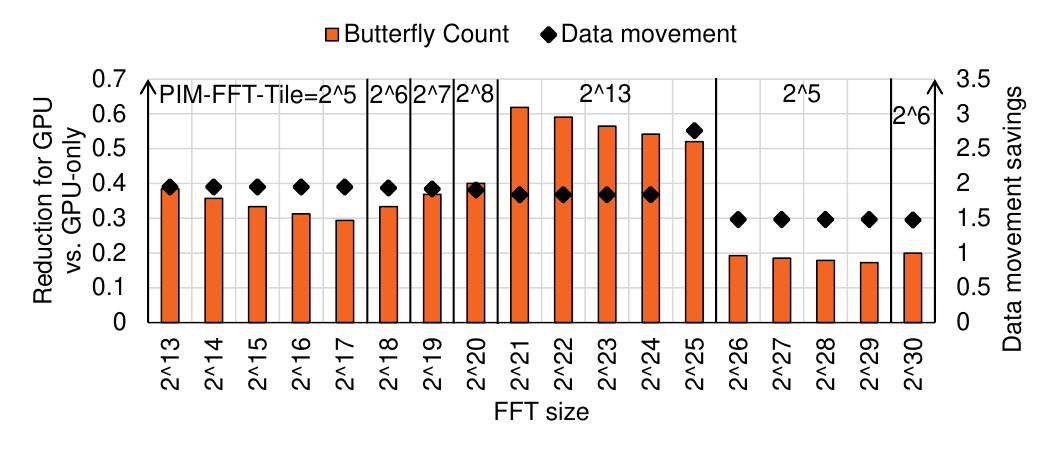}
    \caption{Reduction in overall data movement.}
    \label{fig:overall_host_compute_datamove_reduction}
\end{figure}

We depict the data movement savings of \PNAME in \figref{fig:overall_host_compute_datamove_reduction}. Recall that by offloading portion of the computation to PIM, \PNAME avoids data movement (\revise{GPU} reads/writes from/to HBM) and as such has the potential to lower energy expenditure.\footnote{Note, we account for data movement due to GPU transmitting commands/constants to orchestrate PIM computation} We see this realization in \figref{fig:overall_host_compute_datamove_reduction} as \PNAME leads to 1.48-2.76$\times$ data movement savings for all evaluated FFT sizes (1.81$\times$ on average), which in turn can result in energy savings and therefore improve the overall performance-per-watt. 
We also depict in \figref{fig:overall_host_compute_datamove_reduction} the reduction in FFT computation performed by GPU as a consequence of offloading work to PIM (in terms of butterfly computation count reduction for GPU). Overall, on average, we offload 33\% of the FFT computation to PIM under \PNAME.

\putssec{proposal_opt_sensitivity}{PIM Architecture Sensitivity Studies}

\begin{figure}[t]
    \centering
    \includegraphics[width=\linewidth]{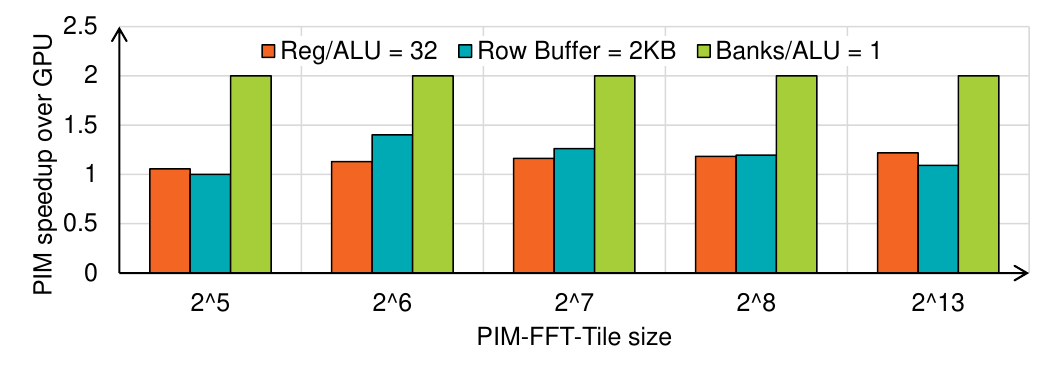}
    \caption{PIM speedup under \PNAME with potential PIM architecture optimizations.} 
    \label{fig:sensitiv_study_tiles}
\end{figure}

We discuss in this section the sensitivity of \PNAME performance to PIM architecture variations. We depict speedups for \textit{PIM-FFT-Tiles} in \figref{fig:sensitiv_study_tiles} and mention \PNAME speedup increase in text (graph not depicted for space reasons).

\noindent\textbf{PIM Register File.}
Similar to register file (RF) usage in baseline GPU, register file size associated with PIM ALU allows reuse of data read from memory. As such, a larger RF size is beneficial as depicted in \figref{fig:sensitiv_study_tiles} where we double the RF size from 16 (baseline) to 32. We see here that doubling the RF size leads to a speedup for \textit{PIM-FFT-Tiles} ranging from  6-22\% which translates to overall \PNAME speedup to be max of 1.41$\times$.

\noindent\textbf{Row-buffer Size.}
Similar to RF size, larger row-buffer (RB) size lowers row-open overheads associated with PIM computation. We study this effect in \figref{fig:sensitiv_study_tiles} by doubling the RB size. While small \textit{PIM-FFT-Tile} sizes like $2^5$ already fit in baseline RB and do not benefit from larger RB size, other \textit{PIM-FFT-Tiles} show speedups (up to 40\% for $2^6$) which translates to overall \PNAME speedup to be max of 1.38$\times$.

\noindent\textbf{PIM Units to Banks Ratio.}
We discuss here the effects of lowering PIM unit sharing between DRAM banks by provisioning a PIM unit per bank. As PIM FFT computation is bottlenecked by compute operations, doubling the number of PIM units, as depicted in \figref{fig:sensitiv_study_tiles} accelerates all the \textit{PIM-FFT-Tiles} by 2$\times$ translating to \PNAME speedup to be max of 1.64$\times$.
\putsec{discuss}{Discussion} 
Herein, we discuss a spectrum of topics of relevance to our work. 

\putssec{discuss_fft_variants}{FFT Variants}
In our work, we focus on a specific FFT variant, namely, radix-2 complex 1D FFT for power-of-two sizes. We discuss here how we can tackle other FFT variants/scenarios. 

\noindent\textbf{Non-2 radix.} Higher radix FFTs (radix-3, radix-5, etc.) are also of interest and can improve compute intensity of FFT computation. While in our work we deduce optimized PIM FFT routine for radix-2, routines for other radixes can also be similarly deduced. 

\noindent\textbf{Non-power-of-two sizes.} Non-power-of-two FFTs are often decomposed. (e.g., as $2^a \times 3^b \times 5^c \times 7^d$ and beyond). While our work already discusses optimized routines for $2^a$ sizes, routines for other blocks can be deduced and judiciously employed.

\noindent\textbf{Real FFTs.} Real input FFTs are also of interest and are typically tackled using complex FFT routines~\cite{msft_fft_paper,msft_fft_paper_sc} we already discuss (e.g., by setting imaginary part of input to zero, packing real inputs into complex input with half the size, etc.). 

\noindent\textbf{Higher-dimension FFTs.} While we discuss 1D FFTs, 2D/3D FFT are also of interest and similarly decomposed to harness on-chip scratchpad space. We can harness our proposed PIM routines to accelerate each dimension separately. 

\noindent\textbf{Precision.} Current PIM prototypes, being focused on machine learning, support 16bit arithmetic with 32bit accumulation. First note that 16bit FFTs on GPUs are also of interest~\cite{tcFFTPaper,cube_nbody,li2020fourier}. If higher-precision FFTs are desired, these can be supported in PIM with additional area expenditure for PIM units. Additionally, since our performance model assumes free compute for GPU, for higher precision (64bit), our speedups will stay intact (PIM compute throughput drop will be matched with GPU memory traffic increase).

\noindent\textbf{Distributed FFT.} We focus in our work on FFT sizes that fit in local memory attached to GPU. For sizes larger than this size, FFT computation is distributed over multiple GPUs. In such cases, PIM can be harnessed for GPU-local portions of computations. In such cases, resultant communication between GPUs can eat into the overall speedup that PIM can provide. That said, accelerating communication is orthogonal to this work. 

\noindent\textbf{Larger scratchpads.} Given how central on-chip scratchpads are for efficient GPU execution, it is possible that scratchpad sizes increase for future GPUs. This will certainly help improve GPU efficiency. However, even in this scenario, for sizes where FFT does not fit in scratchpad, decomposition will be employed, and our proposed PIM variant can be useful for performance and data movement savings. 

\putssec{discuss_pim_software}{PIM Software Implications}
As discussed in \ssecref{proposal_base_data_mapping}, offloading computation to PIM in an efficient manner requires that data be mapped/packed appropriately in memory. Where standalone FFT computations are launched, this can be realized as a one-time cost. Further, for our collaborative decomposition scheme, by prioritizing execution of GPU component, necessary data mapping can be realized by augmenting existing writes from GPU at the end of GPU execution before launching \textit{pim-kernel}. 
\putsec{related}{Related Work}

Given the importance of DFT, FFT is a widely studied primitive and there exists vendor provided FFT libraries for CPUs~\cite{ARMPL,intelMKL,ApplevDSP,AOCLFFTW}, GPUs~\cite{rocfft,cufft,vkfft} and also vendor-independent auto-tuning FFT frameworks such as Fastest Fourier Transform in the West (FFTW)~\cite{fftw98,fftw05}. We believe that our proposed PIM FFT routines can be a good complement to these existing efficient FFT solutions. As an example, \PNAME{} can be integrated in existing FFT libraries as part of the auto-tuning and plan selection process. While this can add complexity to the FFT plan selection, this can be a one-time cost to be reused for a given FFT size. 

Additionally, many prior works which optimize FFT implementation exist such as harnessing built-in generalized matrix multiplication (GEMM) accelerator~\cite{louispishaipdps21,sultandurranipact21,sornahipcw21,tcFFTPaper}, exploiting symmetry and periodicity of the butterflies~\cite{autofftsc19} and more. As accelerators and processors alike are coupled with memory, solutions like PIM can serve as augmentations over and above such existing FFT acceleration solutions that only harness optimizations targeted for the processors. As such, our work complements the rich spectrum of existing (and potentially future) efforts which aim to accelerate the important primitive of FFT.  
\putsec{s07}{Conclusion}

We observe in this work that high-performance implementations of discrete Fourier transforms, aka fast Fourier transform (FFT) are memory bandwidth bound on accelerators such as GPUs. As such, we evaluate in this work the efficacy of emerging commercial processing-in-memory (PIM) solutions, which have a memory bandwidth advantage over GPU by pushing compute to in-memory compute units, to accelerate FFT. By deducing a PIM FFT routine with specialized data mapping and compute orchestration, we see that PIM does not accelerate FFT. To overcome this, we propose collaborative acceleration, which augments existing FFT decomposition mechanism to use our PIM optimized FFT routines. Further, we also proposed hardware augmentation and software optimization to lower PIM operations needed for a given FFT. Our proposed design, \PNAME, which efficiently harnesses PIM, delivers performance of up to 1.38x over a range of FFT sizes and further leads to data movement savings of up to 2.76$\times$.
Overall, our work introduces a complimentary FFT acceleration technique that can be combined with current (and potentially future) processor-side FFT acceleration efforts.


\balance
\bibliographystyle{ACM-Reference-Format}
\bibliography{ref}


\begin{thebibliography}{30}


\ifx \showCODEN    \undefined \def \showCODEN     #1{\unskip}     \fi
\ifx \showDOI      \undefined \def \showDOI       #1{#1}\fi
\ifx \showISBNx    \undefined \def \showISBNx     #1{\unskip}     \fi
\ifx \showISBNxiii \undefined \def \showISBNxiii  #1{\unskip}     \fi
\ifx \showISSN     \undefined \def \showISSN      #1{\unskip}     \fi
\ifx \showLCCN     \undefined \def \showLCCN      #1{\unskip}     \fi
\ifx \shownote     \undefined \def \shownote      #1{#1}          \fi
\ifx \showarticletitle \undefined \def \showarticletitle #1{#1}   \fi
\ifx \showURL      \undefined \def \showURL       {\relax}        \fi
\providecommand\bibfield[2]{#2}
\providecommand\bibinfo[2]{#2}
\providecommand\natexlab[1]{#1}
\providecommand\showeprint[2][]{arXiv:#2}

\bibitem[HBM(2013)]%
        {HBM-jedec}
 \bibinfo{year}{2013}\natexlab{}.
\newblock \bibinfo{title}{JEDEC High Bandwidth Memory (HBM) DRAM}.
\newblock
  \bibinfo{howpublished}{\url{https://www.jedec.org/standards-documents/docs/jesd235a}}.
\newblock


\bibitem[bab(2018)]%
        {babelstream}
 \bibinfo{year}{2018}\natexlab{}.
\newblock \showarticletitle{Evaluating Attainable Memory Bandwidth of Parallel
  Programming Models via BabelStream}.
\newblock \bibinfo{journal}{\emph{Int. J. Comput. Sci. Eng.}}
  \bibinfo{volume}{17}, \bibinfo{number}{3} (\bibinfo{date}{jan}
  \bibinfo{year}{2018}), \bibinfo{pages}{247–262}.
\newblock
\showISSN{1742-7185}


\bibitem[top(2022)]%
        {top500Nov22}
 \bibinfo{year}{2022}\natexlab{}.
\newblock \bibinfo{title}{November 2022 | TOP500}.
\newblock
  \bibinfo{howpublished}{\url{https://www.top500.org/lists/top500/2022/11/}}.
\newblock


\bibitem[mi1(2022)]%
        {mi100PIM}
 \bibinfo{year}{2022}\natexlab{}.
\newblock \bibinfo{title}{Samsung Electronics Semiconductor Unveils
  Cutting-edge Memory Technology to Accelerate Next-generation AI}.
\newblock
  \bibinfo{howpublished}{\url{https://semiconductor.samsung.com/newsroom/tech-blog/samsung-electronics-semiconductor-unveils-cutting-edge-memory-technology-to-accelerate-next-generation-ai/}}.
\newblock


\bibitem[mi2(2023)]%
        {mi210}
 \bibinfo{year}{2023}\natexlab{}.
\newblock \bibinfo{title}{AMD Instinct™ MI210 Accelerator}.
\newblock
  \bibinfo{howpublished}{\url{https://www.amd.com/en/products/server-accelerators/amd-instinct-mi210}}.
\newblock


\bibitem[AOC(2023)]%
        {AOCLFFTW}
 \bibinfo{year}{2023}\natexlab{}.
\newblock \bibinfo{title}{AMD Optimizing CPU Libraries (AOCL) FFTW}.
\newblock
  \bibinfo{howpublished}{\url{https://www.amd.com/en/developer/aocl/fftw.html}}.
\newblock


\bibitem[App(2023)]%
        {ApplevDSP}
 \bibinfo{year}{2023}\natexlab{}.
\newblock \bibinfo{title}{Apple Accelerate Libraries}.
\newblock
  \bibinfo{howpublished}{\url{https://developer.apple.com/documentation/accelerate/vdsp}}.
\newblock


\bibitem[ARM(2023)]%
        {ARMPL}
 \bibinfo{year}{2023}\natexlab{}.
\newblock \bibinfo{title}{Arm Performance Libraries}.
\newblock
  \bibinfo{howpublished}{\url{https://developer.arm.com/downloads/-/arm-performance-libraries}}.
\newblock


\bibitem[bab(2023)]%
        {babelstream_on_amd}
 \bibinfo{year}{2023}\natexlab{}.
\newblock \bibinfo{title}{BabelStream}.
\newblock
  \bibinfo{howpublished}{\url{https://www.amd.com/en/technologies/infinity-hub/babelstream}}.
\newblock


\bibitem[cuf(2023)]%
        {cufft}
 \bibinfo{year}{2023}\natexlab{}.
\newblock \bibinfo{title}{cuFFT}.
\newblock \bibinfo{howpublished}{\url{https://docs.nvidia.com/cuda/cufft/}}.
\newblock


\bibitem[HBM(2023)]%
        {HBM3-jedec}
 \bibinfo{year}{2023}\natexlab{}.
\newblock \bibinfo{title}{HIGH BANDWIDTH MEMORY (HBM3) DRAM}.
\newblock
  \bibinfo{howpublished}{\url{https://www.jedec.org/standards-documents/docs/jesd238a}}.
\newblock


\bibitem[int(2023)]%
        {intelMKL}
 \bibinfo{year}{2023}\natexlab{}.
\newblock \bibinfo{title}{Intel oneAPI Math Kernel Library}.
\newblock
  \bibinfo{howpublished}{\url{https://www.intel.com/content/www/us/en/docs/onemkl/get-started-guide/2023-0/overview.html}}.
\newblock


\bibitem[omn(2023)]%
        {omniperf}
 \bibinfo{year}{2023}\natexlab{}.
\newblock \bibinfo{title}{Omniperf}.
\newblock
  \bibinfo{howpublished}{\url{https://github.com/AMDResearch/omniperf}}.
\newblock


\bibitem[roc(2023)]%
        {rocfft}
 \bibinfo{year}{2023}\natexlab{}.
\newblock \bibinfo{title}{rocFFT}.
\newblock
  \bibinfo{howpublished}{\url{https://github.com/ROCmSoftwarePlatform/rocFFT}}.
\newblock


\bibitem[str(2023)]%
        {stream_cpu}
 \bibinfo{year}{2023}\natexlab{}.
\newblock \bibinfo{title}{STREAM}.
\newblock
  \bibinfo{howpublished}{\url{https://www.cs.virginia.edu/~mccalpin/papers/bandwidth/bandwidth.html}}.
\newblock


\bibitem[Cheng et~al\mbox{.}(2020)]%
        {cube_nbody}
\bibfield{author}{\bibinfo{person}{Shenggan Cheng}, \bibinfo{person}{Hao-Ran
  Yu}, \bibinfo{person}{Derek Inman}, \bibinfo{person}{Qiucheng Liao},
  \bibinfo{person}{Qiaoya Wu}, {and} \bibinfo{person}{James Lin}.}
  \bibinfo{year}{2020}\natexlab{}.
\newblock \showarticletitle{CUBE – Towards an Optimal Scaling of Cosmological
  N-body Simulations}. In \bibinfo{booktitle}{\emph{2020 20th IEEE/ACM
  International Symposium on Cluster, Cloud and Internet Computing (CCGRID)}}.
  \bibinfo{pages}{685--690}.
\newblock
\urldef\tempurl%
\url{https://doi.org/10.1109/CCGrid49817.2020.00-22}
\showDOI{\tempurl}


\bibitem[Cooley and Tukey(1965)]%
        {Cooley1965}
\bibfield{author}{\bibinfo{person}{James~W. Cooley} {and}
  \bibinfo{person}{John~W. Tukey}.} \bibinfo{year}{1965}\natexlab{}.
\newblock \showarticletitle{An algorithm for the machine calculation of complex
  Fourier series}.
\newblock \bibinfo{journal}{\emph{Math. Comp.}}  \bibinfo{volume}{19}
  (\bibinfo{year}{1965}), \bibinfo{pages}{297--301}.
\newblock


\bibitem[Durrani et~al\mbox{.}(2021)]%
        {sultandurranipact21}
\bibfield{author}{\bibinfo{person}{Sultan Durrani},
  \bibinfo{person}{Muhammad~Saad Chughtai}, \bibinfo{person}{Mert Hidayetoglu},
  \bibinfo{person}{Rashid Tahir}, \bibinfo{person}{Abdul Dakkak},
  \bibinfo{person}{Lawrence Rauchwerger}, \bibinfo{person}{Fareed Zaffar},
  {and} \bibinfo{person}{Wen-mei Hwu}.} \bibinfo{year}{2021}\natexlab{}.
\newblock \showarticletitle{Accelerating Fourier and Number Theoretic
  Transforms using Tensor Cores and Warp Shuffles}. In
  \bibinfo{booktitle}{\emph{2021 30th International Conference on Parallel
  Architectures and Compilation Techniques (PACT)}}. \bibinfo{pages}{345--355}.
\newblock
\urldef\tempurl%
\url{https://doi.org/10.1109/PACT52795.2021.00032}
\showDOI{\tempurl}


\bibitem[Frigo and Johnson(1998)]%
        {fftw98}
\bibfield{author}{\bibinfo{person}{M. Frigo} {and} \bibinfo{person}{S.G.
  Johnson}.} \bibinfo{year}{1998}\natexlab{}.
\newblock \showarticletitle{FFTW: an adaptive software architecture for the
  FFT}. In \bibinfo{booktitle}{\emph{Proceedings of the 1998 IEEE International
  Conference on Acoustics, Speech and Signal Processing, ICASSP '98 (Cat.
  No.98CH36181)}}, Vol.~\bibinfo{volume}{3}. \bibinfo{pages}{1381--1384 vol.3}.
\newblock
\urldef\tempurl%
\url{https://doi.org/10.1109/ICASSP.1998.681704}
\showDOI{\tempurl}


\bibitem[Frigo and Johnson(2005)]%
        {fftw05}
\bibfield{author}{\bibinfo{person}{M. Frigo} {and} \bibinfo{person}{S.G.
  Johnson}.} \bibinfo{year}{2005}\natexlab{}.
\newblock \showarticletitle{The Design and Implementation of FFTW3}.
\newblock \bibinfo{journal}{\emph{Proc. IEEE}} \bibinfo{volume}{93},
  \bibinfo{number}{2} (\bibinfo{year}{2005}), \bibinfo{pages}{216--231}.
\newblock
\urldef\tempurl%
\url{https://doi.org/10.1109/JPROC.2004.840301}
\showDOI{\tempurl}


\bibitem[Govindaraju et~al\mbox{.}(2008)]%
        {msft_fft_paper_sc}
\bibfield{author}{\bibinfo{person}{Naga~K. Govindaraju},
  \bibinfo{person}{Brandon Lloyd}, \bibinfo{person}{Yuri Dotsenko},
  \bibinfo{person}{Burton Smith}, {and} \bibinfo{person}{John Manferdelli}.}
  \bibinfo{year}{2008}\natexlab{}.
\newblock \showarticletitle{High performance discrete Fourier transforms on
  graphics processors}. In \bibinfo{booktitle}{\emph{SC '08: Proceedings of the
  2008 ACM/IEEE Conference on Supercomputing}}. \bibinfo{pages}{1--12}.
\newblock
\urldef\tempurl%
\url{https://doi.org/10.1109/SC.2008.5213922}
\showDOI{\tempurl}


\bibitem[Lee et~al\mbox{.}(2021)]%
        {samsungPIM}
\bibfield{author}{\bibinfo{person}{Sukhan Lee}, \bibinfo{person}{Shin-haeng
  Kang}, \bibinfo{person}{Jaehoon Lee}, \bibinfo{person}{Hyeonsu Kim},
  \bibinfo{person}{Eojin Lee}, \bibinfo{person}{Seungwoo Seo},
  \bibinfo{person}{Hosang Yoon}, \bibinfo{person}{Seungwon Lee},
  \bibinfo{person}{Kyounghwan Lim}, \bibinfo{person}{Hyunsung Shin},
  \bibinfo{person}{Jinhyun Kim}, \bibinfo{person}{O Seongil},
  \bibinfo{person}{Anand Iyer}, \bibinfo{person}{David Wang},
  \bibinfo{person}{Kyomin Sohn}, {and} \bibinfo{person}{Nam~Sung Kim}.}
  \bibinfo{year}{2021}\natexlab{}.
\newblock \showarticletitle{Hardware Architecture and Software Stack for PIM
  Based on Commercial DRAM Technology : Industrial Product}. In
  \bibinfo{booktitle}{\emph{2021 ACM/IEEE 48th Annual International Symposium
  on Computer Architecture (ISCA)}}. \bibinfo{pages}{43--56}.
\newblock
\urldef\tempurl%
\url{https://doi.org/10.1109/ISCA52012.2021.00013}
\showDOI{\tempurl}


\bibitem[Lee et~al\mbox{.}(2022)]%
        {hynixPIM}
\bibfield{author}{\bibinfo{person}{Seongju Lee}, \bibinfo{person}{Kyuyoung
  Kim}, \bibinfo{person}{Sanghoon Oh}, \bibinfo{person}{Joonhong Park},
  \bibinfo{person}{Gimoon Hong}, \bibinfo{person}{Dongyoon Ka},
  \bibinfo{person}{Kyudong Hwang}, \bibinfo{person}{Jeongje Park},
  \bibinfo{person}{Kyeongpil Kang}, \bibinfo{person}{Jungyeon Kim},
  \bibinfo{person}{Junyeol Jeon}, \bibinfo{person}{Nahsung Kim},
  \bibinfo{person}{Yongkee Kwon}, \bibinfo{person}{Kornijcuk Vladimir},
  \bibinfo{person}{Woojae Shin}, \bibinfo{person}{Jongsoon Won},
  \bibinfo{person}{Minkyu Lee}, \bibinfo{person}{Hyunha Joo},
  \bibinfo{person}{Haerang Choi}, \bibinfo{person}{Jaewook Lee},
  \bibinfo{person}{Donguc Ko}, \bibinfo{person}{Younggun Jun},
  \bibinfo{person}{Keewon Cho}, \bibinfo{person}{Ilwoong Kim},
  \bibinfo{person}{Choungki Song}, \bibinfo{person}{Chunseok Jeong},
  \bibinfo{person}{Daehan Kwon}, \bibinfo{person}{Jieun Jang},
  \bibinfo{person}{Il Park}, \bibinfo{person}{Junhyun Chun}, {and}
  \bibinfo{person}{Joohwan Cho}.} \bibinfo{year}{2022}\natexlab{}.
\newblock \showarticletitle{A 1ynm 1.25V 8Gb, 16Gb/s/pin GDDR6-based
  Accelerator-in-Memory supporting 1TFLOPS MAC Operation and Various Activation
  Functions for Deep-Learning Applications}. In \bibinfo{booktitle}{\emph{2022
  IEEE International Solid- State Circuits Conference (ISSCC)}},
  Vol.~\bibinfo{volume}{65}. \bibinfo{pages}{1--3}.
\newblock
\urldef\tempurl%
\url{https://doi.org/10.1109/ISSCC42614.2022.9731711}
\showDOI{\tempurl}


\bibitem[Li et~al\mbox{.}(2021)]%
        {tcFFTPaper}
\bibfield{author}{\bibinfo{person}{Binrui Li}, \bibinfo{person}{Shenggan
  Cheng}, {and} \bibinfo{person}{James Lin}.} \bibinfo{year}{2021}\natexlab{}.
\newblock \showarticletitle{tcFFT: A Fast Half-Precision FFT Library for NVIDIA
  Tensor Cores}. In \bibinfo{booktitle}{\emph{2021 IEEE International
  Conference on Cluster Computing (CLUSTER)}}. \bibinfo{pages}{1--11}.
\newblock
\urldef\tempurl%
\url{https://doi.org/10.1109/Cluster48925.2021.00035}
\showDOI{\tempurl}


\bibitem[Li et~al\mbox{.}(2019)]%
        {autofftsc19}
\bibfield{author}{\bibinfo{person}{Zhihao Li}, \bibinfo{person}{Haipeng Jia},
  \bibinfo{person}{Yunquan Zhang}, \bibinfo{person}{Tun Chen},
  \bibinfo{person}{Liang Yuan}, \bibinfo{person}{Luning Cao}, {and}
  \bibinfo{person}{Xiao Wang}.} \bibinfo{year}{2019}\natexlab{}.
\newblock \showarticletitle{AutoFFT: A Template-Based FFT Codes Auto-Generation
  Framework for ARM and X86 CPUs}. In \bibinfo{booktitle}{\emph{Proceedings of
  the International Conference for High Performance Computing, Networking,
  Storage and Analysis}} (Denver, Colorado) \emph{(\bibinfo{series}{SC '19})}.
  \bibinfo{publisher}{Association for Computing Machinery},
  \bibinfo{address}{New York, NY, USA}, Article \bibinfo{articleno}{25},
  \bibinfo{numpages}{15}~pages.
\newblock
\showISBNx{9781450362290}
\urldef\tempurl%
\url{https://doi.org/10.1145/3295500.3356138}
\showDOI{\tempurl}


\bibitem[Li et~al\mbox{.}(2020)]%
        {li2020fourier}
\bibfield{author}{\bibinfo{person}{Zongyi Li}, \bibinfo{person}{Nikola
  Kovachki}, \bibinfo{person}{Kamyar Azizzadenesheli},
  \bibinfo{person}{Burigede Liu}, \bibinfo{person}{Kaushik Bhattacharya},
  \bibinfo{person}{Andrew Stuart}, {and} \bibinfo{person}{Anima Anandkumar}.}
  \bibinfo{year}{2020}\natexlab{}.
\newblock \showarticletitle{Fourier neural operator for parametric partial
  differential equations}.
\newblock \bibinfo{journal}{\emph{arXiv preprint arXiv:2010.08895}}
  (\bibinfo{year}{2020}).
\newblock


\bibitem[Lloyd et~al\mbox{.}(2008)]%
        {msft_fft_paper}
\bibfield{author}{\bibinfo{person}{D.~Brandon Lloyd}, \bibinfo{person}{Chas
  Boyd}, {and} \bibinfo{person}{Naga Govindaraju}.}
  \bibinfo{year}{2008}\natexlab{}.
\newblock \showarticletitle{Fast computation of general Fourier Transforms on
  GPUS}. In \bibinfo{booktitle}{\emph{2008 IEEE International Conference on
  Multimedia and Expo}}. \bibinfo{pages}{5--8}.
\newblock
\urldef\tempurl%
\url{https://doi.org/10.1109/ICME.2008.4607357}
\showDOI{\tempurl}


\bibitem[Pisha and Ligowski(2021)]%
        {louispishaipdps21}
\bibfield{author}{\bibinfo{person}{Louis Pisha} {and} \bibinfo{person}{Lukasz
  Ligowski}.} \bibinfo{year}{2021}\natexlab{}.
\newblock \showarticletitle{Accelerating non-power-of-2 size Fourier transforms
  with GPU Tensor Cores}. In \bibinfo{booktitle}{\emph{2021 IEEE International
  Parallel and Distributed Processing Symposium (IPDPS)}}.
  \bibinfo{pages}{507--516}.
\newblock
\urldef\tempurl%
\url{https://doi.org/10.1109/IPDPS49936.2021.00059}
\showDOI{\tempurl}


\bibitem[Sorna et~al\mbox{.}(2018)]%
        {sornahipcw21}
\bibfield{author}{\bibinfo{person}{Anumeena Sorna}, \bibinfo{person}{Xiaohe
  Cheng}, \bibinfo{person}{Eduardo D'Azevedo}, \bibinfo{person}{Kwai Won},
  {and} \bibinfo{person}{Stanimire Tomov}.} \bibinfo{year}{2018}\natexlab{}.
\newblock \showarticletitle{Optimizing the Fast Fourier Transform Using Mixed
  Precision on Tensor Core Hardware}. In \bibinfo{booktitle}{\emph{2018 IEEE
  25th International Conference on High Performance Computing Workshops
  (HiPCW)}}. \bibinfo{pages}{3--7}.
\newblock
\urldef\tempurl%
\url{https://doi.org/10.1109/HiPCW.2018.8634417}
\showDOI{\tempurl}


\bibitem[Tolmachev(2023)]%
        {vkfft}
\bibfield{author}{\bibinfo{person}{Dmitrii Tolmachev}.}
  \bibinfo{year}{2023}\natexlab{}.
\newblock \showarticletitle{VkFFT-A Performant, Cross-Platform and Open-Source
  GPU FFT Library}.
\newblock \bibinfo{journal}{\emph{IEEE Access}}  \bibinfo{volume}{11}
  (\bibinfo{year}{2023}), \bibinfo{pages}{12039--12058}.
\newblock
\urldef\tempurl%
\url{https://doi.org/10.1109/ACCESS.2023.3242240}
\showDOI{\tempurl}


\end{thebibliography}

\end{document}